\documentclass[journal,10pt]{IEEEtran}

\usepackage{amsmath,amssymb}
\usepackage{subfigure}
\usepackage{graphicx,graphics,color,psfrag}
\usepackage{cite,balance}
\usepackage{algorithm}
\usepackage{accents}
\usepackage{amsthm}
\usepackage{bm}
\usepackage{url}
\usepackage{algorithmic}
\usepackage[english]{babel}
\usepackage{multirow}
\usepackage{enumerate}
\usepackage{cases}
\usepackage{stfloats}
\usepackage{dsfont}
\usepackage{color,soul}
\usepackage{amsfonts}
\usepackage{tcolorbox}

\usepackage{cite,graphicx,amsmath,amssymb}
\usepackage{fancyhdr}
\usepackage{hhline}
\usepackage{graphicx,graphics}
\usepackage{array,color}
\usepackage{amsmath}
\usepackage{amsthm}
\usepackage[flushleft]{threeparttable}
\IEEEoverridecommandlockouts

\newtheorem{proposition}{Proposition}

\include{header}

\newcommand{\mv}[1]{\mbox{\boldmath{$ #1 $}}}

\newcommand{\br}[1]{\bm{\mathrm{#1}}} 
\newcommand{\mr}[1]{\mathrm{#1}}

\makeatletter
\def\endthebibliography{%
	\def\@noitemerr{\@latex@warning{Empty `thebibliography' environment}}%
	\endlist
}
\makeatother

\begin{document}
	\title{Near-Field 3D Localization via MIMO Radar: Cram\'er-Rao Bound Analysis and Estimator Design} 
	\author{Haocheng Hua, \textit{Student Member, IEEE}, Jie Xu, \textit{Senior Member, IEEE}, and Yonina C. Eldar, \textit{Fellow, IEEE} \\
		\thanks{Part of this paper will be presented in IEEE Global Communications Conference (GLOBECOM), Kuala Lumpur, Malaysia, December 4-8, 2023 \cite{hua2023nf}. Haocheng Hua and Jie Xu are with the School of Science and Engineering (SSE) and the Future Network of Intelligence Institute (FNii), The Chinese University of Hong Kong (Shenzhen), Shenzhen, China (e-mail: haochenghua@link.cuhk.edu.cn, xujie@cuhk.edu.cn). J. Xu is the corresponding author.}
		\thanks{Yonina C. Eldar is with the Faculty of Mathematics and Computer
			Science, Weizmann Institute of Science, Rehovot 7610001, Israel (e-mail:
			yonina.eldar@weizmann.ac.il).}
	}
	
	\maketitle
	
	\begin{abstract}
		This paper studies a near-field multiple-input multiple-output (MIMO) radar sensing system, in which the transceivers with massive antennas aim to localize multiple near-field targets in the three-dimensional (3D) space over unknown cluttered environments. We consider a spherical wavefront propagation with both channel phase and amplitude variations over different antennas. 
		Under this setup, the unknown parameters include the 3D coordinates and the complex reflection coefficients of the multiple targets, as well as the noise and interference covariance matrix. First, by considering general transmit signal waveforms, we derive the Fisher information matrix (FIM) corresponding to the 3D coordinates and the complex reflection coefficients of the targets and accordingly obtain the Cram\'er-Rao bound (CRB) for estimating the 3D coordinates. This provides a performance bound for 3D near-field target localization.
		For the special single-target scenario, we obtain the CRB in an analytical form, and analyze its asymptotic scaling behaviors with respect to the target distance and antenna size of the transceiver.
		Next, to facilitate practical localization, we propose two estimators to localize multiple targets based on the maximum likelihood (ML) criterion, namely the 3D approximate cyclic optimization (3D-ACO) and the 3D cyclic optimization with white Gaussian noise (3D-CO-WGN), respectively.
		Numerical results validate the asymptotic CRB analysis and show that the consideration of exact antenna-varying channel amplitudes is essential to achieve accurate CRB and accurate localization in practice when the targets are close to the transceivers.
		It is also shown that the proposed estimators achieve localization performance close to the derived CRB under different cluttered environments, thus validating their effectiveness in practical implementation.
		Furthermore, it is shown that transmit waveforms have a significant impact on CRB and the localization performance.
	\end{abstract}
	
	
	\section{Introduction} 
	\label{sec:intro}
	Sixth-generation (6G) wireless networks have attracted growing research interest in both academia and industry. They are expected to support new usage scenarios such as immersive, massive, and hyper reliable and low-latency communications, as well as integrated sensing and communication (ISAC) \cite{IMT_2030_6G_vision,liu2022integrated, hua2023optimal}. 
	Towards this end, a large number of base stations (BSs) will be densely deployed, which can exploit emerging extremely large-scale multiple-input multiple-output (MIMO) and millimeter wave (mmWave)/terahertz (THz) techniques \cite{heath2016overview} to provide both wireless communications and radar sensing functionalities. 
	Due to the large number of antennas and the high frequency band, the sensing and communications are expected to be implemented in the near-field region (or the Fresnel region \cite{zhang2022beam,zhang20236g,khamidullina2021conditional}), for which the  far-field channel model adopting the planar electromagnetic (EM) wavefront is no longer valid.
	This introduces a paradigm shift from conventional far-field sensing and communications design to new near-field design, for which the spherical wavefront should be considered \cite{lu2021communicating,lu2023near,khamidullina2021conditional,zhang2022beam,cui2022channel,lee1995covariance,grosicki2005weighted,he2021mixed,huang1991near,guerra2021near, yang2023near, zhang20236g, d2022cramer}. For instance, there have been various works investigating extremely large-scale MIMO communications in the near-field by studying signal-to-noise ratio (SNR) scaling laws for single-user communications \cite{lu2021communicating}, beam focusing for multi-user communications \cite{zhang2022beam}, and new channel estimation designs exploiting either the polar-domain sparsity \cite{cui2022channel} or distance-parameterized angular-domain sparsity \cite{zhang2022near}.
	
	
	
	The idea of exploiting the spherical wavefront to localize sensing targets dates back to \cite{huang1991near}, in which the authors studied localization of targets in a two-dimensional (2D) plane by 2D-multiple signal classification (MUSIC) and maximum likelihood estimation (MLE). Since then, various algorithms have been proposed \cite{lee1995covariance,grosicki2005weighted,he2021mixed,yang2023near} for localizing near-field targets in 2D. 
	However, in future wireless sensing scenarios, the target is more likely to be located in three-dimensional (3D) space rather than a 2D plane, particularly when the transceivers are equipped with massive antennas, thus, it is important to consider localizing near-field targets in 3D.
	Towards this end, \cite{khamidullina2021conditional} and \cite{podkurkov2018tensor} compared the performance of several estimators for 3D target localization based on tensor decomposition. The authors in \cite{d2022cramer} studied holographic positioning of a Hertzian dipole source based on MLE. In \cite{guerra2021near}, the authors studied Bayesian tracking algorithms to track a single near-field target in 3D space.
	
	Besides near-field localization algorithm design, another line of research considered the fundamental performance limit of near-field localization by investigating the Cram\'er-Rao bound (CRB), which characterizes the variance lower bound of any unbiased estimator \cite{levy2008principles,song2023intelligent}. Two types of CRBs have been studied in general: the conditional (or deterministic) CRB \cite{el2010conditional,weiss1993range,grosicki2005weighted,gazzah2014crb,khamidullina2021conditional} and the unconditional (or stochastic) CRB \cite{wang2023cram,begriche2012exact,bao2016cramer,el2010conditional,gazzah2014crb,khamidullina2021conditional,sakhnini2022near, d2022cramer}, depending on the assumptions adopted on the underlying signal model. For example, the conditional CRB is applied when the complex reflection coefficients of targets are assumed to be unknown but deterministic while the unconditional CRB is considered when they are assumed to be Gaussian random processes with zero mean and unknown covariance \cite{el2010conditional}. 
	Here, we focus on the conditional CRB of localizing targets in 3D space \cite{khamidullina2021conditional}.
	
	
	
	
	Despite the above progress, prior works have several limitations.
	Most of the previous literature either adopted the Fresnel approximation
	for channel phase characterization \cite{lee1995covariance,bao2016cramer,el2010conditional,grosicki2005weighted} or ignores amplitude variations \cite{podkurkov2018tensor,wang2023cram,begriche2012exact,khamidullina2021conditional,gazzah2014crb,weiss1993range}. Here, we consider the exact spherical wavefront with both channel phase and amplitude variations across antennas for near-field signal modelling \cite{friedlander2019localization}. While a spherical wavefront is adopted in \cite{huang1991near,he2021mixed,sakhnini2022near,guerra2021near,yang2023near,d2022cramer}, they either consider target localization in a 2D plane \cite{huang1991near,he2021mixed,sakhnini2022near,yang2023near} or a single 3D target scenario \cite{guerra2021near,d2022cramer}. 
	In addition, the above works all examine near-field localization solely at the receiver side or the sensor arrays, or treat a bistatic MIMO radar system \cite{podkurkov2018tensor,khamidullina2021conditional,wang2023cram} assuming orthogonal transmit waveforms and matched filtering reception.
	To facilitate transmit waveform optimization for sensing applications in future wireless networks \cite{liu2021cramer, li2007range, hua2022mimojournal}, it is important to enable general transmit waveforms. Finally, localization may face complicated clutter environments in future densely deployed 6G networks. However, all the previous literature assumes that the received signal is only corrupted with white Gaussian noise (WGN), which may not be accurate in unknown
	cluttered environments. 
	Here we derive the CRB and consider estimator design for 3D near-field target localization in MIMO radar systems. We consider generic transmit signal waveforms, exact spherical wavefronts with both channel phase and amplitude variations across antennas, multiple targets, and general modeling of clutter. Under this setup, the unknown parameters at the receivers include the 3D coordinates and the complex reflection coefficients of the targets, as well as the noise and interference covariance matrix. Our goal is to estimate the 3D coordinates of the targets. 
	Our main results are summarized as follows. 
	\begin{itemize} 
		\item First, we derive the Fisher information matrix (FIM) corresponding to the 3D coordinates and the complex reflection coefficients of the targets and accordingly obtain the CRB, which provides a performance bound for 3D near-field target localization.
		As a special case, we provide the closed-form analytical non-matrix expression of the CRB for the single-target scenario and analyze how it scales asymptotically with respect to the target distance and antenna size of the transceiver. 
		It is shown that when the target is close to the transceiver, our considered model with varying channel amplitudes achieves more accurate CRB and more accurate localization in practice than that without such consideration. 
		\item Next, we propose two practical estimators to localize the multiple targets based on the maximum likelihood (ML) criterion: 3D approximate cyclic optimization (3D-ACO) and 3D cyclic optimization with white Gaussian noise (3D-CO-WGN). The 3D-ACO is applied for any unknown cluttered environment without presuming the structure of the noise and interference covariance matrix, while the 3D-CO-WGN is derived assuming that the noise at the receiver is WGN.
		%
		\item Finally, we conduct numerical experiments to validate the CRB analysis and evaluate the performance of the proposed estimators. In the scenario with WGN, the  3D-CO-WGN is shown to outperform 3D-ACO, and both of them approach the CRB in the high SNR regime. In the non-WGN scenario, the 3D-ACO performs better than 3D-CO-WGN. It is also shown that with proper non-isotropic transmit waveform, the achieved CRB and localization  performance by practical estimators can outperform those by the conventional design with isotropic transmission, thus showing the significance of considering the generic transmit waveforms.  These results demonstrate the effectiveness of the proposed estimators in practical scenarios and the potential of transmit waveform adaptation.
	\end{itemize}
	
	{\it Notations:} Boldface letters refer to vectors (lower case) or matrices (upper case). For a square matrix $\mv{M}$, ${\operatorname{tr}}(\mv{M})$, $\bm{M}^{-1}$, and $|\bm{M}|$ denote its trace, inverse, and determinant, respectively. For an arbitrary-size matrix $\mv{M}$, $\mathfrak{R}(\bm{M})$, $\mathfrak{I}(\bm{M})$, $\bm{M}^H$, $\bm{M}^*$, $\bm{M}^T$, $\bm{M}[m:n,p:q]$, and $\operatorname{vec}(\bm{M})$ denote its real part, imaginary part, conjugate transpose, conjugate, transpose, the corresponding sub-block matrix with dimension $(n-m+1) \times (q-p+1)$, and its vectorization, respectively, and $\odot$ is the Hadamard product.
	The distribution of a circularly symmetric complex Gaussian (CSCG) random vector with mean vector $\mv{x}$ and covariance matrix $\mv{\Sigma}$ is denoted by $\mathcal{CN}(\mv{x,\Sigma})$; and $\sim$ stands for ``distributed as''. We use $\mathbb{R}^{x\times y}$ and $\mathbb{C}^{x\times y}$ to represent the spaces of real and complex matrices with dimension $x \times y$, respectively. {${\mathbb{E}}\{\cdot\}$} denotes the statistical expectation. $\|\mv{x}\|$ is the Euclidean norm of a complex vector $\mv{x}$ and $\operatorname{diag}(\bm{x})$ denotes a diagonal matrix with diagonal elements $\bm{x}$. The imaginary unit is written as $\mathrm{j} = \sqrt{-1}$ .

	\section{System Model}\label{Section_system_model}
	
\begin{figure}[t]
		\centering
		\includegraphics[width=3.3in]{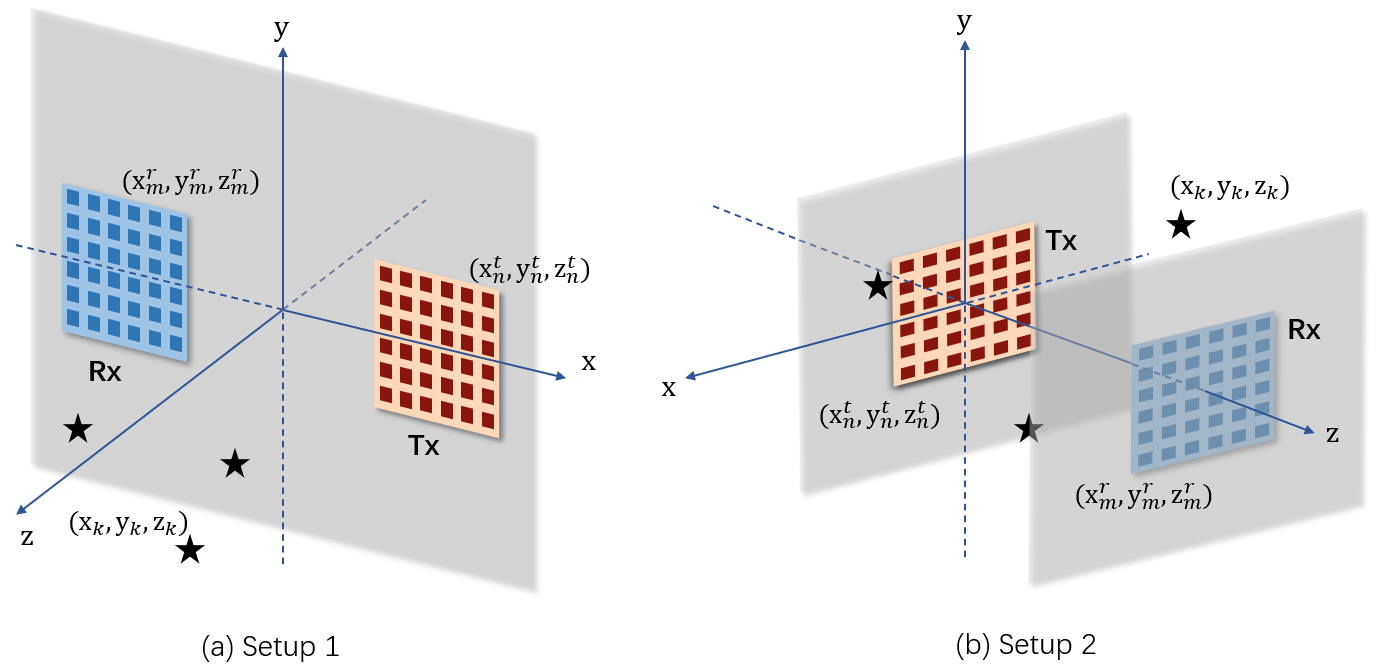}
		\centering
		\caption{MIMO radar operating in the near-field region. Black stars denote the targets to be localized: (a) Antenna arrays at Rx and Tx are located in the same plane; (b) Antenna arrays at Rx and Tx face towards each other.}
		\label{fig:sm}
	\end{figure}

	We consider a MIMO radar system as shown in Fig. \ref{fig:sm}, which consists of a radar transmitter (Tx) with $N$ antennas and a radar receiver (Rx) with $M$ antennas. The Tx and Rx antenna arrays can be deployed in the same or different planes, as shown in Fig. \ref{fig:sm}(a) and \ref{fig:sm}(b), respectively. Note that  Fig. \ref{fig:sm}(a) may contain the monostatic MIMO radar as a special case when  the Tx and Rx overlapp.
	Let $\br{l}^t_{n}  = (\mr{x}^t_n,\mathrm{y}^t_n,\mathrm{z}^t_n), n \in \mathcal{N} \triangleq \{1,2,...,N\}$, and $\bm{\mathrm{l}}^r_{m} = (\mr{x}^r_m,\mathrm{y}^r_m,\mathrm{z}^r_m), m \in \mathcal{M} \triangleq \{1,2,...,M\}$, denote the position of the $n$-th antenna at Tx and of the $m$-th antenna at Rx, respectively. We assume there exists $K$ sensing targets located in the near-field region of the MIMO radar. Let $\mathcal{K} \triangleq \{1,...,K\}$ denote the set of targets and $\bm{\mathrm{l}}_k = (\mathrm{x}_k,\mathrm{y}_k,\mathrm{z}_k), k \in \mathcal{K}$ the position of the $k$-th target. 
	
	Let $\bm{x}_l$ denote the transmit signal at the Tx in symbol $l \in \{1,...,L\}$ for multi-target localization
	and $\bm{X} = \left[\bm{x}_1,...,\bm{x}_L\right] \in \mathbb{C}^{N \times L}$ represent the overall transmit signals over the $L$ symbols in one radar dwell time. The sample covariance matrix of the transmitted signals is then given by
	\begin{align}\label{eq:remain_def_Fisher_Rx}
		\bm{R}_X = \frac{1}{L} \bm{X} \bm{X}^H.
	\end{align}
	Under a narrowband assumption, the received signal over the $L$ symbols is described by a matrix $\bm{Y} \in \mathbb{C}^{M \times L}$, given by
	\begin{align}\label{equ:Rx_data_matrix}
		\bm{Y} = \sum_{k = 1}^{K} b_k \bm{a}(\bm{\mathrm{l}}_k)  \bm{v}^T(\bm{\mathrm{l}}_k) \bm{X} + \bm{Z},
	\end{align}
	where $\bm{Z} = \left[\bm{z}_1,...,\bm{z}_L\right] \in \mathbb{C}^{M \times L}$ denotes the noise and interference with each column being independent and identically distributed (i.i.d.) CSCG random vectors with zero mean and covariance $\bm{Q} \in \mathbb{C}^{M \times M}$, and $b_k$ denotes the target complex reflection coefficients proportional to the radar-cross-sections (RCS) of the $k$-th target. 
	Here, we consider unknown cluttered environments, so that $\bm{Q}$ is unknown. Furthermore, in (\ref{equ:Rx_data_matrix}),
	$\bm{a}(\bm{\mathrm{l}}_k)$ and $\bm{v}(\bm{\mathrm{l}}_k)$ denote the steering vectors at the Rx and Tx for the $k$-th target, respectively, and are given by 
	\begin{align}\label{equ:steer_Rx}
		\bm{a}(\bm{\mathrm{l}}_k) = [\alpha_1^r(\bm{\mathrm{l}}_k)  e^{-\mathrm{j} \nu \|\bm{\mathrm{l}}^r_1 - \bm{\mathrm{l}}_{k}\|},..., \alpha_{M}^r(\bm{\mathrm{l}}_k) e^{-\mathrm{j} \nu \|\bm{\mathrm{l}}^r_M - \bm{\mathrm{l}}_{k}\|} ]^T,
	\end{align}
	\begin{align}\label{equ:steer_Tx}
		\bm{v}(\bm{\mathrm{l}}_k) = [\alpha_1^t(\bm{\mathrm{l}}_k)e^{-\mathrm{j} \nu \|\bm{\mathrm{l}}^t_1 - \bm{\mathrm{l}}_{k}\|},..., \alpha_N^t(\bm{\mathrm{l}}_k)e^{-\mathrm{j} \nu \|\bm{\mathrm{l}}^t_N - \bm{\mathrm{l}}_{k}\|} ]^T,
	\end{align}
	where $\nu = \frac{2 \pi}{\lambda}$ is the wave number of the carrier, $\lambda$ is the wavelength, $\alpha_m^r(\bm{\mathrm{l}}_k) = \frac{\lambda}{4 \pi \|\bm{\mathrm{l}}_k - \bm{\mathrm{l}}^r_m\|}$ denotes the distance-dependent channel amplitude from the $k$-th target to the $m$-th receive antenna based on the free space path-loss model, and $\alpha_n^t(\bm{\mathrm{l}}_k) = \frac{\lambda}{4 \pi \|\bm{\mathrm{l}}_k - \bm{\mathrm{l}}^t_{n}\|}$ denotes that from the $k$-th target to the $n$-th transmit antenna. Note that in (\ref{equ:steer_Rx}) and (\ref{equ:steer_Tx}), we assume omnidirectional antennas with unit antenna gains.
	
	For ease of exposition, the received signal $\bm{Y}$ in (\ref{equ:Rx_data_matrix}) is expressed in a more compact form as
	\begin{align}\label{equ:data_compact_rx}
		\bm{Y} = \bm{A}(\{\bm{\mathrm{l}}_k\}) \bm{B} \bm{V}^T(\{\bm{\mathrm{l}}_k\}) \bm{X} + \bm{Z},
	\end{align}
	with
	\begin{align}
		\label{equ:A_compact}
		& \bm{A}(\{\bm{\mathrm{l}}_k\}) = \left[\bm{a}(\bm{\mathrm{l}}_1),\bm{a}(\bm{\mathrm{l}}_2),...,\bm{a}(\bm{\mathrm{l}}_K)\right] \in \mathbb{C}^{M \times K}, \\
		\label{equ:V_compact}
		& \bm{V}(\{\bm{\mathrm{l}}_k\})  = \left[\bm{v}(\bm{\mathrm{l}}_1),\bm{v}(\bm{\mathrm{l}}_2),...,\bm{v}(\bm{\mathrm{l}}_K)\right] \in \mathbb{C}^{N \times K}, \\
		\label{equ:b_complete_RI}
		& \bm{b}  = \left[b_1,...,b_K\right]^T =  \left[b_{\text{R}_1}+ \mathrm{j}b_{\text{I}_1},...,b_{\text{R}_K}+\mathrm{j}b_{\text{I}_K}\right]^T, \\
		\label{equ:b_compact}
		& \qquad \qquad \qquad \bm{B}  = \operatorname{diag}(\bm{b}),
	\end{align}
	where $b_{\text{R}_k}$ and $b_{\text{I}_k}$ denote the real and imaginary parts of $b_k$, respectively. 
	In (\ref{equ:data_compact_rx}), the unknown parameters include the 3D target locations $\{\bm{\mathrm{l}}_k\}$, the complex reflection coefficients $\bm{b}$, and the noise and interference covariance matrix $\bm{Q}$. 
	
	\section{Near-Field CRB}\label{sec:nf_CRB}
	
	In this section, we derive the CRB for estimating the 3D coordinates of the multiple near-field targets based on the received signal $\bm{Y}$ in (\ref{equ:data_compact_rx}). In particular, we first derive the CRB for the general case with multiple targets in Section \ref{subsec:general_CRB}
	and then provide a closed-form analytical non-matrix expression of the CRB for localizing a single target with the corresponding asymptotic analysis in Section \ref{subsec:closed_form_CRB_single}. 
	
	\subsection{General Case with Multiple Targets}\label{subsec:general_CRB}
	Let $\tilde{\bm{\theta}} \in \mathbb{R}^{5K+M^2}$ denote a vector containing all the real unknowns in the target-related parameter vector $\bm{\theta}$, defined as
	\begin{align}\label{eq:theta_interested}
		\nonumber
		\bm{\theta} & = \left[\mathrm{x}_1,...,\mathrm{x}_K,\mathrm{y}_1,...,\mathrm{y}_K,\mathrm{z}_1,...,\mathrm{z}_K,b_{\text{R}_1},..,b_{\text{R}_K},b_{\text{I}_1},...,b_{\text{I}_K}\right]^T \\
		& = \left[\bm{\mathrm{x}},\bm{\mathrm{y}},\bm{\mathrm{z}},\bm{b}_\text{R},\bm{b}_\text{I}\right]^T \in \mathbb{R}^{5K},
	\end{align}
	and the $M^2$ unknown nuisance real parameters in $\bm{Q}$. Let $\bm{y} \triangleq \operatorname{vec}(\bm{Y})$. It follows from (\ref{equ:data_compact_rx}) that
	\begin{align}\label{eq:vec_Y}
		\bm{y} = \left[(\bm{A} \bm{B} \bm{V}^T \bm{x}_1)^T+\bm{z}_1^T,...,(\bm{A} \bm{B} \bm{V}^T \bm{x}_L)^T+\bm{z}_L^T\right]^T,
	\end{align}
	which is a complex Gaussian random vector with mean vector
	\begin{align}\label{equ:Rx_data_mat_DGC}
		\bm{\mathrm{\mu}}(\tilde{\bm{\theta}}) = \left[(\bm{A} \bm{B} \bm{V}^T \bm{x}_1)^T,...,(\bm{A} \bm{B} \bm{V}^T \bm{x}_L)^T\right]^T
	\end{align}
	and covariance matrix
	\begin{align}\label{equ:cov_C}
		\bm{\bm{C}}(\tilde{\bm{\theta}}) = \left[\begin{array}{ccc}
			\bm{Q} & \bm{0} & \bm{0} \\
			\bm{0}&  \ddots & \bm{0}\\
			\bm{0}  &\bm{0} & \bm{Q}  
		\end{array}\right] \in \mathbb{C}^{ML \times ML},
	\end{align}
	i.e., $\bm{y} \sim \mathcal{CN}(\bm{\mu}(\tilde{\bm{\theta}}),\bm{C}(\tilde{\bm{\theta}}))$. Here, (\ref{equ:cov_C}) holds since the observations at different snapshots are uncorrelated with each other. 
	We then have the following proposition.
	\begin{proposition}\label{Pro:overall_fisher}
		\emph{Based on (\ref{eq:vec_Y}), the overall FIM of $\tilde{\bm{\theta}}$, denoted as $\tilde{\bm{\mathrm{F}}} \in \mathbb{R}^{(5K+M^2) \times (5K+M^2)}$, is given as
			\begin{align}\label{eq:overall_fisher}
				\tilde{\bm{\mathrm{F}}} & (\tilde{\bm{\theta}}_i,\tilde{\bm{\theta}}_j)   = L \operatorname{tr}\left[\bm{Q}^{-1} \frac{\partial \bm{Q}}{\partial \tilde{\bm{\theta}}_i} \bm{Q}^{-1} \frac{\partial \bm{Q}}{\partial \tilde{\bm{\theta}}_j}\right] + \\
				\nonumber
				& 2 \mathfrak{R} \operatorname{tr} \left[ (\frac{\partial (\bm{A} \bm{B} \bm{V}^T \bm{X})}{\partial \tilde{\bm{\theta}}_i})^H \bm{Q}^{-1} (\frac{\partial (\bm{A} \bm{B} \bm{V}^T \bm{X})}{\partial \tilde{\bm{\theta}}_j}) \right],
			\end{align}
			where $\tilde{\bm{\mathrm{F}}}(\tilde{\bm{\theta}}_i,\tilde{\bm{\theta}}_j)$ denotes the $(i,j)$-th element of the FIM $\tilde{\bm{\mathrm{F}}}$.
			\begin{proof}
				See Appendix \ref{Appendix:Proof_Prop_1}.
		\end{proof}}
	\end{proposition}

	Based on Proposition \ref{Pro:overall_fisher}, $\tilde{\bm{\mathrm{F}}}\left[1:5K,5K+1:5K+M^2\right] = \bm{0}_{5K \times M^2}$. This is due to the fact that in each element of $\tilde{\bm{\mathrm{F}}} (\tilde{\bm{\theta}}_i,\tilde{\bm{\theta}}_j)$, $\tilde{\bm{\theta}}_i \in \bm{\theta}$ and $\tilde{\bm{\theta}}_j \in \tilde{\bm{\theta}} \setminus \bm{\theta}$, with $\setminus$ denoting the difference of two sets, and as a result, the derivatives in (\ref{eq:overall_fisher}) will vanish. Accordingly, it holds that $\tilde{\bm{\mathrm{F}}}\left[5K+1:5K+M^2,1:5K\right] = \tilde{\bm{\mathrm{F}}}^T\left[1:5K,5K+1:5K+M^2\right] = \bm{0}_{M^2 \times 5K}$. Thus, the overall FIM $\tilde{\bm{\mathrm{F}}}$ is a block diagonal matrix with respect to the target-related unknowns in $\bm{\theta}$ and the nuisance parameters in $\bm{Q}$. Therefore, we can calculate the CRBs of $\bm{\theta}$ and $\bm{Q}$ separately. As we are mainly interested in the CRBs of $\{\bm{\mathrm{l}}_k\}$, we just need to find the FIM $\bm{\mathrm{F}}$ corresponding to $\{\{\bm{\mathrm{l}}_k\}, \bm{b}\}$, i.e., $\bm{\mathrm{F}} = \tilde{\bm{\mathrm{F}}}[1:5K,1:5K]$, which is shown in the following.

	\begin{proposition}\label{Pro:F_deri}
		\emph{The FIM $\bm{\mathrm{F}} \in \mathbb{R}^{5K \times 5K}$ with respect to $\bm{\theta}$ is
			\begin{align}\label{eq:fisher_mat_sim}
				2 \left[
				\arraycolsep=0.75pt\def\arraystretch{1.2}
				\begin{array}{ccccc}
					\mathfrak{R}(\bm{\mathrm{F}}_{\bm{\mathrm{xx}}}) & \mathfrak{R}(\bm{\mathrm{F}}_{\bm{\mathrm{xy}}}) & \mathfrak{R}(\bm{\mathrm{F}}_{\bm{\mathrm{xz}}}) & \mathfrak{R}(\bm{\mathrm{F}}_{\bm{\mathrm{xb}}}) & -\mathfrak{I}(\bm{\mathrm{F}}_{\bm{\mathrm{xb}}}) \\[1pt]
					\mathfrak{R}(\bm{\mathrm{F}}^T_{\bm{\mathrm{xy}}}) & \mathfrak{R}(\bm{\mathrm{F}}_{\bm{\mathrm{yy}}}) & \mathfrak{R}(\bm{\mathrm{F}}_{\bm{\mathrm{yz}}}) & \mathfrak{R}(\bm{\mathrm{F}}_{\bm{\mathrm{yb}}}) & -\mathfrak{I}(\bm{\mathrm{F}}_{\bm{\mathrm{yb}}}) \\[1pt]
					\mathfrak{R}(\bm{\mathrm{F}}^T_{\bm{\mathrm{xz}}}) & \mathfrak{R}(\bm{\mathrm{F}}^T_{\bm{\mathrm{yz}}}) & \mathfrak{R}(\bm{\mathrm{F}}_{\bm{\mathrm{zz}}}) & \mathfrak{R}(\bm{\mathrm{F}}_{\bm{\mathrm{zb}}}) & -\mathfrak{I}(\bm{\mathrm{F}}_{\bm{\mathrm{zb}}}) \\[1pt]
					\mathfrak{R}(\bm{\mathrm{F}}^T_{\bm{\mathrm{xb}}}) & \mathfrak{R}(\bm{\mathrm{F}}^T_{\bm{\mathrm{yb}}}) & \mathfrak{R}(\bm{\mathrm{F}}^T_{\bm{\mathrm{zb}}}) & \mathfrak{R}(\bm{\mathrm{F}}_{\bm{\mathrm{bb}}}) & -\mathfrak{I}(\bm{\mathrm{F}}_{\bm{\mathrm{bb}}}) \\[1pt]
					-\mathfrak{I}(\bm{\mathrm{F}}^T_{\bm{\mathrm{xb}}}) & -\mathfrak{I}(\bm{\mathrm{F}}^T_{\bm{\mathrm{yb}}}) & -\mathfrak{I}(\bm{\mathrm{F}}^T_{\bm{\mathrm{zb}}}) & -\mathfrak{I}(\bm{\mathrm{F}}^T_{\bm{\mathrm{bb}}}) & \mathfrak{R}(\bm{\mathrm{F}}_{\bm{\mathrm{bb}}})
				\end{array}\right], 
			\end{align}
			where $\bm{\mathrm{F}}_{\bm{\mathrm{xx}}}, \bm{\mathrm{F}}_{\bm{\mathrm{yy}}}$, and $\bm{\mathrm{F}}_{\bm{\mathrm{zz}}}$ are given by
			\begin{align}\label{eq:F_multiple_list_xx_yy_zz}
				\bm{\mathrm{F}}_{\bm{\mathrm{uu}}} & =  L (\dot{\bm{A}}_{\bm{\mathrm{u}}}^H  \bm{Q}^{-1} \dot{\bm{A}_{\bm{\mathrm{u}}}}) \odot (\bm{B}^* \bm{V}^H \bm{R}_X^* \bm{V} \bm{B}) \\
				\nonumber
				& + L (\dot{\bm{A}}_{\bm{\mathrm{u}}}^H  \bm{Q}^{-1} \bm{A}) \odot (\bm{B}^* \bm{V}^H \bm{R}_X^* \dot{\bm{V}}_{\bm{\mathrm{u}}} \bm{B}) \\
				\nonumber
				& + L (\bm{A}^H  \bm{Q}^{-1} \dot{\bm{A}_{\bm{\mathrm{u}}}}) \odot (\bm{B}^* \dot{\bm{V}}_{\bm{\mathrm{u}}}^H \bm{R}_X^* \bm{V} \bm{B})\\
				\nonumber
				& + L (\bm{A}^H  \bm{Q}^{-1} \bm{A}) \odot (\bm{B}^* \dot{\bm{V}}_{\bm{\mathrm{u}}}^H \bm{R}_X^* \dot{\bm{V}}_{\bm{\mathrm{u}}} \bm{B}), \enspace \bm{\mathrm{u}} \in \{\bm{\mathrm{x}},\bm{\mathrm{y}},\bm{\mathrm{z}}\}, 
			\end{align}
			$\bm{\mathrm{F}}_{\bm{\mathrm{xy}}}, \bm{\mathrm{F}}_{\bm{\mathrm{xz}}}$, and $\bm{\mathrm{F}}_{\bm{\mathrm{yz}}}$ are given by
			\begin{align}\label{eq:F_multiple_list_xy_xz_yz}
				& \bm{\mathrm{F}}_{\bm{\mathrm{uv}}}  =  L (\dot{\bm{A}}_{\bm{\mathrm{u}}}^H  \bm{Q}^{-1} \dot{\bm{A}_{\bm{\mathrm{v}}}}) \odot (\bm{B}^* \bm{V}^H \bm{R}_X^* \bm{V} \bm{B}) \\
				\nonumber
				& + L (\dot{\bm{A}}_{\bm{\mathrm{u}}}^H  \bm{Q}^{-1} \bm{A}) \odot (\bm{B}^* \bm{V}^H \bm{R}_X^* \dot{\bm{V}}_{\bm{\mathrm{v}}} \bm{B}) \\
				\nonumber
				& + L (\bm{A}^H  \bm{Q}^{-1} \dot{\bm{A}_{\bm{\mathrm{v}}}}) \odot (\bm{B}^* \dot{\bm{V}}_{\bm{\mathrm{u}}}^H \bm{R}_X^* \bm{V} \bm{B})\\
				\nonumber
				& + L (\bm{A}^H  \bm{Q}^{-1} \bm{A}) \odot (\bm{B}^* \dot{\bm{V}}_{\bm{\mathrm{u}}}^H \bm{R}_X^* \dot{\bm{V}}_{\bm{\mathrm{v}}} \bm{B}), \enspace \bm{\mathrm{uv}} \in \{\bm{\mathrm{xy}},\bm{\mathrm{xz}},\bm{\mathrm{yz}}\}, 
			\end{align}
			and $\bm{\mathrm{F}}_{\bm{\mathrm{bb}}}$, $\bm{\mathrm{F}}_{\bm{\mathrm{xb}}}, \bm{\mathrm{F}}_{\bm{\mathrm{yb}}}$, and $\bm{\mathrm{F}}_{\bm{\mathrm{zb}}}$ are given by
			\begin{align}\label{eq:F_multiple_list_xb_yb_zb}
				\bm{\mathrm{F}}_{\bm{\mathrm{bb}}} & = L (\bm{A}^H  \bm{Q}^{-1} \bm{A}) \odot (\bm{V}^H \bm{R}_X^* \bm{V}),\\
				\bm{\mathrm{F}}_{\bm{\mathrm{ub}}}  & =  L (\dot{\bm{A}}_{\bm{\mathrm{u}}}^H  \bm{Q}^{-1} \bm{A}) \odot (\bm{B}^* \bm{V}^H \bm{R}_X^* \bm{V}) \\
				\nonumber
				& + L (\bm{A}^H  \bm{Q}^{-1} \bm{A}) \odot (\bm{B}^* \dot{\bm{V}}_{\bm{\mathrm{u}}}^H \bm{R}_X^* \bm{V}), \enspace \bm{\mathrm{u}} \in \{\bm{\mathrm{x}},\bm{\mathrm{y}},\bm{\mathrm{z}}\}, 
			\end{align}
			respectively. Here,
			\begin{align}\label{eq:VX_partial}
				\dot{\bm{A}_{\bm{\mathrm{u}}}} & = \left[\frac{\partial \bm{a}(\bm{\mathrm{l}}_1)}{\partial \mathrm{u}_1},...,\frac{\partial \bm{a}(\bm{\mathrm{l}}_K)}{\partial \mathrm{u}_K}\right], \enspace \bm{\mathrm{u}} \in \{\bm{\mathrm{x}},\bm{\mathrm{y}},\bm{\mathrm{z}}\},   \\
				\dot{\bm{V}_{\bm{\mathrm{u}}}} & = \left[\frac{\partial \bm{v}(\bm{\mathrm{l}}_1)}{\partial \mathrm{u}_1},...,\frac{\partial \bm{v}(\bm{\mathrm{l}}_K)}{\partial \mathrm{u}_K}\right], \enspace \bm{\mathrm{u}} \in \{\bm{\mathrm{x}},\bm{\mathrm{y}},\bm{\mathrm{z}}\}, 
			\end{align}
			\begin{align}
				\label{eq:AX_partial_element}
				\left(\frac{\partial \bm{a}(\bm{\mathrm{l}}_k)}{\partial \mathrm{u}_k}\right)_m & = \bm{a}_m(\bm{\mathrm{l}}_k)(\frac{\mathrm{u}_m^r-\mathrm{u}_k}{\|\bm{\mathrm{l}}^r_m - \bm{\mathrm{l}}_{k}\|^2}+\mathrm{j} \nu \frac{\mathrm{u}_m^r-\mathrm{u}_k}{\|\bm{\mathrm{l}}^r_m - \bm{\mathrm{l}}_{k}\|}),   \\
				\label{eq:VX_partial_element}
				\left(\frac{\partial \bm{v}(\bm{\mathrm{l}}_k)}{\partial \mathrm{u}_k}\right)_n & = \bm{v}_n(\bm{\mathrm{l}}_k)(\frac{\mathrm{u}_n^t-\mathrm{u}_k}{\|\bm{\mathrm{l}}^t_n - \bm{\mathrm{l}}_{k}\|^2}+\mathrm{j} \nu \frac{\mathrm{u}_n^t-\mathrm{u}_k}{\|\bm{\mathrm{l}}^t_n - \bm{\mathrm{l}}_{k}\|}). 
			\end{align}
			In (\ref{eq:AX_partial_element}) and (\ref{eq:VX_partial_element}), $\bm{a}_m(\bm{\mathrm{l}}_k)$ and $\bm{v}_n(\bm{\mathrm{l}}_k)$ denote the $m$-th element of $\bm{a}(\bm{\mathrm{l}}_k)$ and the $n$-th element of $\bm{v}(\bm{\mathrm{l}}_k)$, respectively.}
		\begin{proof}
			See Appendix \ref{Appendix:Proof_Prop_F_deri}.
		\end{proof}
	\end{proposition}

	%
	
	Using Proposition \ref{Pro:F_deri}, we have the complete FIM $\bm{\mathrm{F}}$ corresponding to the target-related parameters in $\bm{\theta}$. The CRB matrix $\bm{\mathrm{C}}$ for estimating $\bm{\theta}$ is given as
	\begin{align}\label{eq:CRB_mat}
		\bm{\mathrm{C}} = \bm{\mathrm{F}}^{-1}.
	\end{align}
	According to (\ref{eq:theta_interested}),  we obtain the sum CRB for estimating the position of the $k$-th target $\bm{\mathrm{l}}_k$ as
	\begin{align}\label{eq:CRB_position}
		\text{CRB}_k & =  \text{CRB}_{k,\mathrm{x}} + \text{CRB}_{k,\mathrm{y}} + \text{CRB}_{k,\mathrm{z}}\\
		\nonumber
		& = \bm{\mathrm{C}}[k,k] + \bm{\mathrm{C}}[k+K,k+K] + \bm{\mathrm{C}}[k+2K,k+2K],
	\end{align}
	where $\text{CRB}_{k,\mathrm{u}}, \mathrm{u} \in \{\mathrm{x},\mathrm{y},\mathrm{z}\}$ denotes the CRB for estimating the $\mathrm{u}$-coordinate of the $k$-th target.
	
	
	The CRB derived here is more general than that in \cite{khamidullina2021conditional} in the following aspects. First, in (\ref{eq:remain_def_Fisher_Rx}) and (\ref{equ:Rx_data_matrix}) we consider general transmit waveforms with sample covariance matrix $\bm{R}_X$ being any arbitrary positive semi-definite matrix, while \cite{khamidullina2021conditional} focused on orthogonal waveforms with $\bm{R}_X = \bm{I}$. Next, in (\ref{equ:Rx_data_matrix}) we consider general unknown noise and interference covariance matrix $\bm{Q}$, while \cite{khamidullina2021conditional} assumed WGN model with $\bm{Q} = \sigma^2  \bm{I}$. Finally, we consider general channel amplitude variations across antennas in (\ref{equ:steer_Rx}) and (\ref{equ:steer_Tx}), while \cite{khamidullina2021conditional} treated a simplified constant-amplitude model with $\alpha_1^r(\bm{\mathrm{l}}_k)=...=\alpha_M^r(\bm{\mathrm{l}}_k) \triangleq \alpha^r(\bm{\mathrm{l}}_k)$ and $\alpha_1^t(\bm{\mathrm{l}}_k)=...=\alpha_N^t(\bm{\mathrm{l}}_k) \triangleq \alpha^t(\bm{\mathrm{l}}_k)$, where the round-trip path loss are included in the target complex reflection coefficients.


	\subsection{Special Case with One Single Target and WGN}
	\label{subsec:closed_form_CRB_single}
	
	In this subsection, we derive the closed-form expression of the CRB in the special case when there exists only one sensing target with WGN. In this case, $\bm{Q} = \sigma^2 \bm{I}$, where $\sigma^2$ is the noise power at each receive antenna to be estimated. We then have the following proposition.
	\begin{proposition}\label{prop:CRB_single_WGN}
		\emph{The CRB of the position of a single target under WGN at Rx is given as
			\begin{align}\label{eq:CRB_p_sum_xyz}
				\text{CRB} = \text{CRB}_\mathrm{x} + \text{CRB}_\mathrm{y} + \text{CRB}_\mathrm{z}, 
			\end{align} 
			with
			\begin{align}
				\nonumber
				\text{CRB}_\mathrm{x} & = \frac{\sigma^2 \left[(\tilde{f}_{\mathrm{y}\mathrm{y}}-f_{\mathrm{y}\mathrm{y}}^R)(\tilde{f}_{\mathrm{z}\mathrm{z}}-f_{\mathrm{z}\mathrm{z}}^R)-(\mathfrak{R}(\tilde{f}_{\mathrm{y}\mathrm{z}}-f_{\mathrm{y}\mathrm{z}}^R))^2\right]}{2 |b|^2 L |\tilde{\bm{D}}|}, \\
				\nonumber
				\text{CRB}_\mathrm{y} & = \frac{\sigma^2 \left[(\tilde{f}_{\mathrm{x}\mathrm{x}}-f_{\mathrm{x}\mathrm{x}}^R)(\tilde{f}_{\mathrm{z}\mathrm{z}}-f_{\mathrm{z}\mathrm{z}}^R)-(\mathfrak{R}(\tilde{f}_{\mathrm{x}\mathrm{z}}-f_{\mathrm{x}\mathrm{z}}^R))^2\right]}{2 |b|^2 L |\tilde{\bm{D}}|}, \\
				\nonumber
				\text{CRB}_\mathrm{z} & = \frac{\sigma^2 \left[(\tilde{f}_{\mathrm{x}\mathrm{x}}-f_{\mathrm{x}\mathrm{x}}^R)(\tilde{f}_{\mathrm{y}\mathrm{y}}-f_{\mathrm{y}\mathrm{y}}^R)-(\mathfrak{R}(\tilde{f}_{\mathrm{x}\mathrm{y}}-f_{\mathrm{x}\mathrm{y}}^R))^2\right]}{2 |b|^2 L |\tilde{\bm{D}}|},
			\end{align} 
			where 
			\begin{align}\label{eq:D_tilde}
				\tilde{\bm{D}} = \left[
				\arraycolsep=2.9pt\def\arraystretch{0.9}
				\begin{array}{lll}
					\tilde{f}_{\mathrm{x}\mathrm{x}}-f_{\mathrm{x}\mathrm{x}}^R & \mathfrak{R}(\tilde{f}_{\mathrm{x}\mathrm{y}}-f_{\mathrm{x}\mathrm{y}}^R) & \mathfrak{R}(\tilde{f}_{\mathrm{x}\mathrm{z}}-f_{\mathrm{x}\mathrm{z}}^R) \\
					\mathfrak{R}(\tilde{f}_{\mathrm{x}\mathrm{y}}-f_{\mathrm{x}\mathrm{y}}^R) & \tilde{f}_{\mathrm{y}\mathrm{y}}-f_{\mathrm{y}\mathrm{y}}^R & \mathfrak{R}(\tilde{f}_{\mathrm{y}\mathrm{z}}-f_{\mathrm{y}\mathrm{z}}^R) \\
					\mathfrak{R}(\tilde{f}_{\mathrm{x}\mathrm{z}}-f_{\mathrm{x}\mathrm{z}}^R) & \mathfrak{R}(\tilde{f}_{\mathrm{y}\mathrm{z}}-f_{\mathrm{y}\mathrm{z}}^R) & \tilde{f}_{\mathrm{z}\mathrm{z}}-f_{\mathrm{z}\mathrm{z}}^R
				\end{array}\right],
			\end{align}
			\begin{align}
				\nonumber
				\tilde{f}_{\mathrm{u}\mathrm{u}}   =  & \|\dot{\bm{a}}_\mathrm{u}\|^2 \bm{v}^H \bm{R}_X^* \bm{v} + \dot{\bm{a}}_\mathrm{u}^H \bm{a} \bm{v}^H \bm{R}_X^* \dot{\bm{v}}_\mathrm{u} + \bm{a}^H \dot{\bm{a}}_\mathrm{u} \dot{\bm{v}}_\mathrm{u}^H \bm{R}_X^* \bm{v} \\
				\label{eq:f_uu}
				&  + \|\bm{a}\|^2 \dot{\bm{v}}_\mathrm{u}^H \bm{R}_X^* \dot{\bm{v}}_\mathrm{u}, \enspace \mathrm{u} \in \{\mathrm{x},\mathrm{y},\mathrm{z}\}, \\
				\nonumber
				\tilde{f}_{\mathrm{u}\mathrm{v}}  = & \dot{\bm{a}}_\mathrm{u}^H \dot{\bm{a}}_\mathrm{v} \bm{v}^H \bm{R}_X^* \bm{v} + \dot{\bm{a}}_\mathrm{u}^H \bm{a} \bm{v}^H \bm{R}_X^* \dot{\bm{v}}_\mathrm{v} + \bm{a}^H \dot{\bm{a}}_\mathrm{v} \dot{\bm{v}}_\mathrm{u}^H \bm{R}_X^* \bm{v} \\
				&  + \|\bm{a}\|^2 \dot{\bm{v}}_\mathrm{u}^H \bm{R}_X^* \dot{\bm{v}}_\mathrm{v}, \enspace \mathrm{u}\mathrm{v} \in \{\mathrm{x}\mathrm{y},\mathrm{x}\mathrm{z},\mathrm{y}\mathrm{z}\}, 
			\end{align}
			\begin{align}
				& f_{\mathrm{u}\mathrm{u}}^R  =  \frac{|\dot{\bm{a}}_\mathrm{u}^H \bm{a} \bm{v}^H \bm{R}_X^* \bm{v} + \|\bm{a}\|^2 \dot{\bm{v}}_\mathrm{u}^H \bm{R}_X^* \bm{v} |^2}{\|\bm{a}\|^2 \bm{v}^H \bm{R}_X^* \bm{v}}, \mathrm{u} \in \{\mathrm{x},\mathrm{y},\mathrm{z}\},  \\
				\nonumber
				& f_{\mathrm{u}\mathrm{v}}^R  =  \frac{1}{\|\bm{a}\|^2 \bm{v}^H \bm{R}_X^* \bm{v}}  (\dot{\bm{a}}_\mathrm{u}^H \bm{a} \bm{v}^H \bm{R}_X^* \bm{v} + \|\bm{a}\|^2 \dot{\bm{v}}_\mathrm{u}^H \bm{R}_X^* \bm{v})\\
				& (\dot{\bm{a}}_\mathrm{v}^H \bm{a} \bm{v}^H \bm{R}_X^* \bm{v} + \|\bm{a}\|^2 \dot{\bm{v}}_\mathrm{v}^H \bm{R}_X^* \bm{v})^H, \enspace \mathrm{u}\mathrm{v} \in \{\mathrm{x}\mathrm{y},\mathrm{x}\mathrm{z},\mathrm{y}\mathrm{z}\}.
				\label{eq:f_yz_R}
			\end{align}
		}
		\begin{proof}
			See Appendix \ref{Appendix:proof_CRB_single_WGN}.
		\end{proof}
	\end{proposition}
	
	It is easy to see that the CRB of the target position is a function of the sample covariance matrix of the transmitted waveform $\bm{R}_X$. It is also observed that the CRB is proportional to the noise power $\sigma^2$ and inversely proportional to the square of the amplitude of the complex reflection coefficient $|b|^2$ and the length of the total transmitted waveform $L$. To gain more insight, in the following we analyze how CRB scales and behaves with respect to the target distance and antenna sizes of the transceiver.
	
\begin{figure}[t]
		\centering
		\includegraphics[width=1.7in]{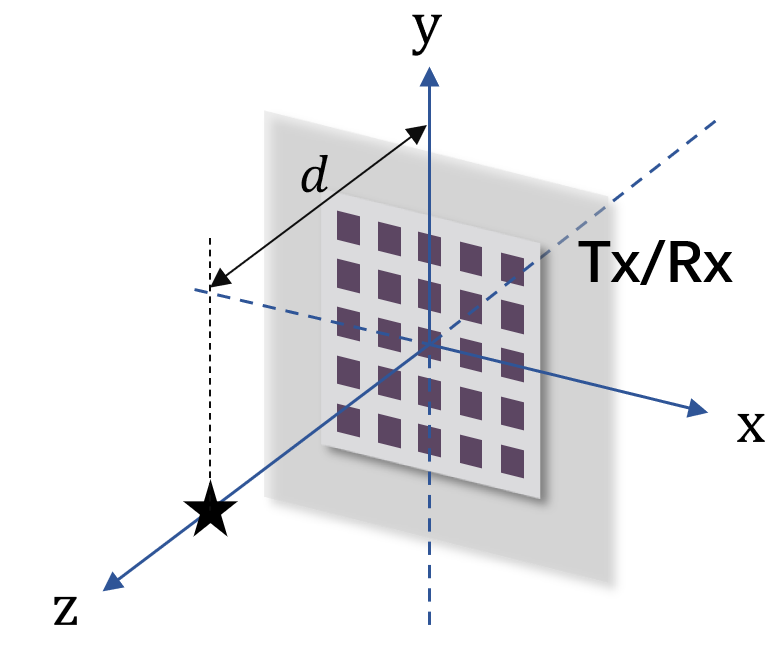}
		\centering
		\caption{The simplified scenario where there is only one target marked as a black star, with position $(0,0,d)$. }
	\label{fig:sm_sim}
\end{figure}

	Towards this end, we further consider a special monostatic MIMO radar setup in Fig. \ref{fig:sm_sim} with a single target. 
	Under this setup, Rx/Tx is completely overlapping  while the target is assumed to lie on the $\mathrm{z}$-axis with position $(0,0,d)$.
	Without loss of generality, throughout the remaining part of this subsection, we consider a square uniform planar array (UPA), with $N = M = n^2$, where $n$ is an odd number and represents the number of antennas in both $\mathrm{x}$ and $\mathrm{y}$ dimension. Let $s$ be the uniform spacing between adjacent antennas. The center of the arrays is assumed to coincide with the $\mathrm{z}$-axis. 
	Besides, the commonly used transmission strategy, i.e., isotropic transmission with $\bm{R}_X = \bm{I}$, is adopted.
	Based on the above setup, we aim to express $\text{CRB}$ in (\ref{eq:CRB_p_sum_xyz}) in terms of the target distance from the center of the planar array $d$, the antenna spacing $s$, and the number of antennas in each dimension $n$ and accordingly analyze its asymptotic behavior. We have the following proposition.
\begin{proposition}\label{prop:CRB_sim_expression}
	\emph{Under the setup in Fig. \ref{fig:sm_sim}, the closed-form CRB of the position of the target is given as
		\begin{align}\label{eq:CRB_sim_expression}
			\text{CRB} = \frac{\sigma^2}{4 |b|^2 L} \left[\frac{2}{\|\bm{a}\|^2 \|\dot{\bm{a}}_\mathrm{x}\|^2} + \frac{1}{ \|\bm{a}\|^2 \|\dot{\bm{a}}_\mathrm{z}\|^2 - |\dot{\bm{a}}_\mathrm{z}^H \bm{a}|^2 }\right],
		\end{align}
		where 
		\begin{align}\label{eq:CRB_sim_expression_xy}
			\text{CRB}_\mathrm{x} = \text{CRB}_\mathrm{y} = \frac{\sigma^2}{4 |b|^2 L} \frac{1}{\|\bm{a}\|^2 \|\dot{\bm{a}}_\mathrm{x}\|^2},
		\end{align}
		\begin{align}\label{eq:CRB_sim_expression_z}
			\text{CRB}_\mathrm{z} = \frac{\sigma^2}{4 |b|^2 L} \frac{1}{ \|\bm{a}\|^2 \|\dot{\bm{a}}_\mathrm{z}\|^2 - |\dot{\bm{a}}_\mathrm{z}^H \bm{a}|^2 },
		\end{align}
		with
		\begin{align}
			\label{eq:a_norm_pds}
			& \|\bm{a}\|^2   = \frac{1}{4 \nu^2} \left(\frac{1}{d^2} + 4 \sum_{i=0}^{\frac{n-1}{2}} \sum_{k=1}^{\frac{n-1}{2}} \frac{1}{d^2 + i^2s^2 + k^2 s^2} \right), \\
			&  \|\dot{\bm{a}}_\mathrm{x}\|^2 =  \frac{D_1^\mathrm{x}+ \nu^2 D_2^\mathrm{x}}{4 \nu^2}, \enspace \|\dot{\bm{a}}_\mathrm{z}\|^2 = \frac{d^2}{4 \nu^2} (D_1^\mathrm{z}+\nu^2 D_2^\mathrm{z}), \\
			\label{eq:a_x_der}
			& \dot{\bm{a}}_\mathrm{z}^H \bm{a} = \frac{d}{4 \nu^2} (- D_2^\mathrm{z} + \mathrm{j} \nu D_3^\mathrm{z}), 
		\end{align}
		where
		\begin{align}
			\nonumber
			D_1^\mathrm{x} = 
			4 \sum_{i=1}^{\frac{n-1}{2}} \sum_{k=1}^{\frac{n-1}{2}} \frac{k^2 s^2}{(d^2 + i^2s^2 + k^2 s^2)^3} + 2  \sum_{k=1}^{\frac{n-1}{2}} \frac{k^2 s^2}{(d^2 + k^2 s^2)^3}, 
		\end{align}
		\begin{align}
			\nonumber
			D_2^\mathrm{x} = 4 \sum_{i=1}^{\frac{n-1}{2}} \sum_{k=1}^{\frac{n-1}{2}} \frac{k^2 s^2}{(d^2 + i^2s^2 + k^2 s^2)^2} + 2  \sum_{k=1}^{\frac{n-1}{2}} \frac{k^2 s^2}{(d^2 + k^2 s^2)^2},
		\end{align}
		\begin{align}
			\nonumber
			& D_1^\mathrm{z} = \frac{1}{d^6} + 4 \sum_{i=0}^{\frac{n-1}{2}} \sum_{k=1}^{\frac{n-1}{2}} \frac{1}{(d^2 + i^2s^2 + k^2 s^2)^3}, \\
			\nonumber
			& D_2^\mathrm{z} = \frac{1}{d^4} + 4 \sum_{i=0}^{\frac{n-1}{2}} \sum_{k=1}^{\frac{n-1}{2}} \frac{1}{(d^2 + i^2s^2 + k^2 s^2)^2},
		\end{align}
		\begin{align}
			\nonumber
			& D_3^\mathrm{z} = \frac{1}{d^3} + 4 \sum_{i=0}^{\frac{n-1}{2}} \sum_{k=1}^{\frac{n-1}{2}} \frac{1}{(d^2 + i^2s^2 + k^2 s^2)^{\frac{3}{2}}},
		\end{align}	
	}
	\begin{proof}
		See Appendix \ref{Appendix:proof_CRB_sim_expression}.
	\end{proof}
\end{proposition}



The CRB in Proposition \ref{prop:CRB_sim_expression} is expressed as a function of three key system parameters, i.e., $\text{CRB} = f(n,d,s)$. 

\begin{proposition}\label{prop:asy_CRB_d_large}
	\emph{Under the setup in Fig. \ref{fig:sm_sim}, we have
		\begin{align}\label{eq:CRB_xy_asymp}
			\lim\limits_{d \to \infty} \frac{\text{CRB}_\mathrm{x}}{d^6} = \lim\limits_{d \to \infty} \frac{\text{CRB}_\mathrm{y}}{d^6} =  \frac{48 \sigma^2}{|b|^2 L}  \frac{\nu^2}{(n^2-1)n^4s^2},
		\end{align}
		and 
		\begin{align}\label{eq:CRB_z_asymp}
			\lim\limits_{d \to \infty} \frac{\text{CRB}_\mathrm{z}}{d^8} = \frac{1440 \sigma^2}{|b|^2 L} \frac{\nu^2}{(n^2-1)n^4(n^2-4)s^4}.
		\end{align}
	}
	\begin{proof}
		See Appendix \ref{Appendix:proof_asy_CRB_d_large}.
	\end{proof}
\end{proposition}
According to Proposition \ref{prop:asy_CRB_d_large}, under the monostatic MIMO radar setup in Fig. \ref{fig:sm_sim}, when $d$ is sufficiently large, $\text{CRB}_\mathrm{x}$, $\text{CRB}_\mathrm{y}$, and $\text{CRB}_\mathrm{z}$ are approximated by
\begin{align}\label{eq:CRB_x_appro_d_large}
	\text{CRB}_\mathrm{x} = \text{CRB}_\mathrm{y} \approx \frac{48 \sigma^2}{|b|^2 L}  \frac{\nu^2}{(n^2-1)n^4s^2} d^6,
\end{align}	
and
\begin{align}\label{eq:CRB_z_appro_d_large}
	\text{CRB}_\mathrm{z} \approx \frac{1440 \sigma^2}{|b|^2 L} \frac{\nu^2}{(n^2-1)n^4(n^2-4)s^4} d^8.
\end{align}
When $d$ increases, the CRB of $\mathrm{z}$-coordinate deteriorates much faster than the CRB of $\mathrm{x}$-coordinate and $\mathrm{y}$-coordinate. This is very intuitive, as the resolution of $\mathrm{z}$-coordinate in this special case corresponds to the resolution of the distance between the target and the array. When the target is moving further away, the spherical wavefront gradually becomes planar and the array is not able to resolve the distance accurately any more. It is also worth noticing that $\text{CRB}_\mathrm{z}$ is inversely proportional to $n^8$, which is in sharp contrast to $n^6$ for $\text{CRB}_\mathrm{x}$ and $\text{CRB}_\mathrm{y}$. This again validates the importance of large antenna arrays in localizing the exact 3D coordinates of the targets. 

When $d$ is fixed, and the antenna aperture or $n$ increases, we have the following proposition.

\begin{proposition}\label{prop:asy_CRB_P_large}
	\emph{Under the setup in Fig. \ref{fig:sm_sim}, we have
		\begin{align}\label{eq:CRB_xy_P_inf}
			& \lim\limits_{n \to \infty}  \text{CRB}_\mathrm{x}  = \lim\limits_{n \to \infty} \text{CRB}_\mathrm{y} = 0,\\
			\label{eq:CRB_z_P_inf}
			& \lim\limits_{n \to \infty} \text{CRB}_\mathrm{z} = 0.
		\end{align}
		Accordingly, $\lim\limits_{n \to \infty} \text{CRB} = 0$.
	}
	\begin{proof}
		See Appendix \ref{Appendix:proof_asy_CRB_P_large}.
	\end{proof}
\end{proposition}

	With fixed distance $d$ between the target and the center of the antenna array and enlarging antenna aperture, it is  indispensable to adopt an exact spherical wavefront model considering both distance-dependent variations in both channel phases and amplitudes over different antennas.	Under this scenario, Proposition \ref{prop:asy_CRB_P_large} shows that larger antenna aperture always helps reduce the total CRB of the target position and when $n \to \infty$, the CRB will asymptotically approach zero. 


\section{Practical Estimators Design} \label{Section_estimators}



We next present two practical near-field localization algorithms based on (\ref{equ:data_compact_rx}) following the ML criterion, namely 3D-ACO and 3D-CO-WGN, respectively.

\subsection{3D-ACO}\label{subsec:alg_3D_ACO}
The 3D-ACO estimator is composed of two main steps. The first is to obtain the concentrated negative log-likelihood function solely based on the 3D coordinates of all targets. The second is to apply cyclic optimization technique \cite{li2008mimo} to resolve the 3D coordinate of each target.

We first discuss how to obtain the concentrated negative log-likelihood function. 
Recall that  $\bm{y} \triangleq \operatorname{vec}(\bm{Y}) \sim \mathcal{CN}(\bm{\mu},\bm{C})$, with $\bm{\mu}$ and $\bm{C}$ given in (\ref{equ:Rx_data_mat_DGC}) and (\ref{equ:cov_C}), respectively. Here, we omit $\tilde{\bm{\theta}}$ for simplicity. The negative log-likelihood function of $\bm{y}$ is then expressed as
\begin{align}\label{eq:initial_neg_log_like}
	- \ln f(\bm{y}) = -\ln \frac{1}{\pi^{L M} |\bm{C}|} + (\bm{y}-\bm{\mu})^H \bm{C}^{-1} (\bm{y}-\bm{\mu}).
\end{align}
With the constant term omitted, the negative log-likelihood in (\ref{eq:initial_neg_log_like}) is re-expressed as
\begin{align}\label{eq:neg_log_like}
	\nonumber
	f_1(\bm{Q},\{b_k\},\{\bm{\mathrm{l}}_k\}) & = L \ln |\bm{Q}| + \operatorname{tr}\left[(\bm{Y} - \bm{A} \bm{B} \bm{V}^T \bm{X}) \right. \\
	& \left. (\bm{Y} - \bm{A} \bm{B} \bm{V}^T \bm{X})^H \bm{Q}^{-1}\right].
\end{align}
Let $\bm{W} \triangleq (\bm{Y} - \bm{A} \bm{B} \bm{V}^T \bm{X})(\bm{Y} - \bm{A} \bm{B} \bm{V}^T \bm{X})^H$.
We then have
\begin{align}\label{eq:f_1_expression}
	f_1(\bm{Q},\{b_k\},\{\bm{\mathrm{l}}_k\}) = L \ln |\bm{Q}| + \operatorname{tr}(\bm{W} \bm{Q}^{-1}).
\end{align}

By maximizing the log-likelihood or equivalently minimizing $f_1$ with respect to $\bm{Q}$, we obtain the estimate of $\bm{Q}$ as 
\begin{align}\label{eq:Est_Q_opt}
	\bm{Q}^\star = \frac{1}{L} \bm{W}.
\end{align}
Substituting (\ref{eq:Est_Q_opt}) into (\ref{eq:f_1_expression}), we have
\begin{align}\label{eq:f_1_with_W}
	f_1 (\frac{1}{L} \bm{W}, \{b_k\}, \{\bm{\mathrm{l}}_k\}) = L \ln (\frac{1}{L})^M + L \ln |\bm{W}| + L M.
\end{align}
Ignoring the first and the third constant terms, the minimization of $f_1$ in (\ref{eq:f_1_with_W}) is equivalent to minimizing
\begin{align}\label{equ:f2_AML}
	& f_2(\{b_k\}, \{\bm{\mathrm{l}}_k\})   = L \ln |\bm{W}| \\
	\nonumber
	& = L \ln |(\bm{Y} - \bm{A} \operatorname{diag}(\bm{b}) \bm{V}^T \bm{X})(\bm{Y} - \bm{A} \operatorname{diag}(\bm{b}) \bm{V}^T \bm{X})^H|.
\end{align}

Next, we seek to minimize $f_2$ with respect to 3D target locations $\{\bm{\mathrm{l}}_k\}$ and target reflection coefficients $\bm{b}$. We first obtain the estimation of the complex reflection coefficient vector $\bm{b}$ with given $\{\bm{\mathrm{l}}_k\}$. Notice that the received data matrix in (\ref{equ:data_compact_rx}) is in canonical form known as the diagonal growth curve (DGC) model \cite{xu2006diagonal}. Therefore, we apply the approximate maximum likelihood (AML) estimator in \cite{xu2006diagonal} to obtain an AML estimate of $\bm{b}$ as
\begin{align}\label{equ:AML_Est_b}
	\bm{b}  =  \left[(\bm{A}^H \bm{\mathrm{J}}^{-1} \bm{A}) \odot (\bm{\mathrm{S}} \bm{\mathrm{S}}^H)^T \right]^{-1} \operatorname{vecd}\left(\bm{A}^H \bm{\mathrm{J}}^{-1} \bm{Y} \bm{\mathrm{S}}^H\right), 
\end{align}
where 
$\operatorname{vecd}(\cdot)$ denotes a column vector formed by the diagonal elements of a given matrix, and 
\begin{align}
	\label{equ:def_S}
	\bm{\mathrm{S}} & = \bm{V}^T \bm{X}, \\
	\label{equ:def_J}
	\bm{\mathrm{J}} & = \frac{1}{L} \bm{Y} \bm{Y}^H - \frac{1}{L} \bm{Y} \bm{\mathrm{S}}^H (\bm{\mathrm{S}} \bm{\mathrm{S}}^H)^{-1} \bm{\mathrm{S}} \bm{Y}^H.
\end{align}
Combining 
(\ref{equ:f2_AML})-(\ref{equ:def_J}), 
we obtain the concentrated negative log-likelihood function $f_3(\{\bm{\mathrm{l}}_k\})$ as
\begin{align}\label{equ:f1_neg_log}
	f_3(\{\bm{\mathrm{l}}_k\}) = L \ln |(\bm{Y} - \bm{A} \operatorname{diag}(\bm{b}) \bm{\mathrm{S}})(\bm{Y} - \bm{A} \operatorname{diag}(\bm{b}) \bm{\mathrm{S}})^H|,
\end{align}
with $\bm{b}$, $\bm{\mathrm{S}}$, and $\bm{\mathrm{J}}$ given in (\ref{equ:AML_Est_b}), (\ref{equ:def_S}), and (\ref{equ:def_J}), respectively.
In (\ref{equ:f1_neg_log}), the number of unknowns is reduced to $3 K$. 

We now proceed to resolve the locations of all the targets, i.e., $\{\bm{\mathrm{l}}_k\}$ by minimizing $f_3(\{\bm{\mathrm{l}}_k\})$. Towards this end, we apply the cyclic optimization technique, which is summarized in Algorithm 1 and is implemented in a cyclic manner. To start with, we first estimate the location of a single target by finding the minimizer of $f_3$ in (\ref{equ:f1_neg_log}) via a 3D exhaustive search (Steps 1-3 in Algorithm 1). Then, if $K_{\text{max}}>1$, we gradually increase the expected number of targets $\hat{K}$. Under given $\hat{K}$, we estimate the locations of the $\hat{K}$ targets cyclically. In other words, in each iteration, we estimate the location of each target as the one minimizing $f_3$ in (\ref{equ:f1_neg_log}) via a 3D exhaustive search, in which the estimated locations of the other targets are considered fixed (Steps 6-19 in Algorithm 1). The cyclic iteration terminates until some convergence criterion is met, e.g., the relative change of $f_3$ between two consecutive iterations is less than some pre-determined threshold $\epsilon$ (Step 11 in Algorithm 1). Finally, the outer iteration terminates when the number of expected targets $\hat{K}$ reaches a prescribed maximum number $K_{\text{max}}$ (Step 4 in Algorithm 1).


\begin{algorithm}[t]
	\label{alg:3D_CO}
	\textbf{Algorithm 1} 3D-ACO for targets localization
	\hrule
	\begin{itemize}
		\item \textbf{Input}: $\bm{X}$, $\bm{Y}$, $K_{\text{max}}$, $\epsilon$, and $\bm{\mathrm{L}} \in \mathbb{R}^{3 \times K_{\text{max}}}$
		
		\begin{algorithmic}[1]
			\STATE $\hat{K} \gets 1$ 
			\STATE Estimate $\bm{\mathrm{l}}_1 = \left[\mathrm{x}_1,\mathrm{y}_1,\mathrm{z}_1\right]^T$ as the one minimizing $f_3$ in (\ref{equ:f1_neg_log})
			\STATE $\bm{\mathrm{L}}\left[:,1\right] \gets \bm{\mathrm{l}}_1$ 
			\WHILE  {$\hat{K} < K_{\text{max}}$}
			\STATE $\hat{K} \gets \hat{K}+1$
			\STATE Estimate $\bm{\mathrm{l}}_{\hat{K}} = \left[\mathrm{x}_{\hat{K}},\mathrm{y}_{\hat{K}},\mathrm{z}_{\hat{K}}\right]^T$ as the one minimizing $f_3$ in (\ref{equ:f1_neg_log}) with fixed $\{\bm{\mathrm{l}}_m\}_{m=1}^{\hat{K}-1}$ 
			\STATE Update $f_3$ based on  $\{\bm{\mathrm{l}}_m\}_{m=1}^{\hat{K}}$ with newly estimated $\bm{\mathrm{l}}_{\hat{K}}$
			\STATE $\bm{\mathrm{L}}\left[:,\hat{K}\right] \gets \bm{\mathrm{l}}_{\hat{K}}$ 
			\STATE $f_\text{old} \gets  f_3(\{\bm{\mathrm{l}}_m\}_{m=1}^{\hat{K}}) + 2\epsilon, \text{ } f_\text{new} \gets f_3(\{\bm{\mathrm{l}}_m\}_{m=1}^{\hat{K}})$
			\STATE $  p \gets 1$
			\WHILE {$\left(f_\text{old} - f_\text{new} > \epsilon\right)$} 
		\STATE $ f_\text{old} \gets f_\text{new}$
		\STATE Estimate $\bm{\mathrm{l}}_p = \left[\mathrm{x}_p,\mathrm{y}_p,\mathrm{z}_p\right]^T$ as the one minimizing $f_3$ in (\ref{equ:f1_neg_log}) with fixed $\{\bm{\mathrm{l}}_m\}_{m=1,m \neq p}^{\hat{K}}$ 
		\STATE Update $f_3$ based on $\{\bm{\mathrm{l}}_m\}_{m=1}^{\hat{K}}$ with newly estimated $\bm{\mathrm{l}}_p$
		\STATE $\bm{\mathrm{L}}\left[:,p\right] \gets \bm{\mathrm{l}}_p$ 
		\STATE $f_\text{new} \gets f_3(\{\bm{\mathrm{l}}_m\}_{m=1}^{\hat{K}})$
		\STATE $p \gets \mod (p,\hat{K})+1$
		\ENDWHILE
		\ENDWHILE
		\STATE $\bm{\mathrm{l}}_k^{\text{ACO}} = \bm{\mathrm{L}}\left[:,k\right], k = 1,2,..., K_\text{max}$
	\end{algorithmic}
	\item \textbf{Output}: $\{\bm{\mathrm{l}}_k^{\text{ACO}}\}_{k=1}^{K_{\text{max}}}$
\end{itemize}
\end{algorithm}

The complexity of each 3D exhaustive search is typically large. To reduce the complexity,
we first specify the initial search grid 
and obtain a coarse estimate of the location of the target. Then, based on the obtained coarse estimation, we iteratively find better estimation around the previous estimate with reduced grid search size in each iteration.

Note that we assume the number of the targets is known as $K$ in advance and accordingly set it as the maximum searching number $K_{\text{max}}$ in Algorithm 1. The estimator can be extended to the case when $K$ is not  \textit{a-priori} known by combining it with the model order selection rules \cite{stoica2004model}. 

\subsection{3D-CO-WGN Estimator under WGN}

The proposed 3D-ACO estimator introduced in the previous subsection can be applied to general $\bm{Q}$. However, if prior knowledge about the structure of $\bm{Q}$ is available in advance, we can find a better estimator. Assume
\begin{align}\label{eq:specific_wgn_Q}
\bm{Q} = \sigma^2 \bm{I},
\end{align}  
where $\sigma^2$ is the unknown noise variance and is assumed to be the same across all the receive antennas, we derive another estimator based on this widely adopted assumption of $\bm{Q}$ in the following.

With $\bm{Q}$ specified in (\ref{eq:specific_wgn_Q}), the negative likelihood function in (\ref{eq:neg_log_like}) is simplified into
\begin{align}\label{eq:neg_log_like_wgn}
\nonumber
f_1^{\text{WGN}}(\sigma^2,\{b_k\},&\{\bm{\mathrm{l}}_k\})   = L M \ln \sigma^2 + \\
& \frac{1}{\sigma^2} \sum_{l=1}^L \|\bm{y}_l - \bm{A} \operatorname{diag}(\bm{b}) \bm{V}^T \bm{x}_l\|^2,
\end{align}
where $\sigma^2$ is unknown. Taking the derivative of $f_1^{\text{WGN}}$ with respect to $\sigma^2$ and setting it equal to zero, we have
\begin{align}\label{equ:est_sigma_2}
(\sigma^2)^\star = \frac{1}{L M} \sum_{l=1}^L \|\bm{y}_l - \bm{A} \operatorname{diag}(\bm{b}) \bm{V}^T \bm{x}_l\|^2.
\end{align}
Plugging this estimate of $\sigma^2$ back into (\ref{eq:neg_log_like_wgn}), we have
\begin{align}\label{equ:neg_likelihood_with_sigma2}
\nonumber
f_1^{\text{WGN}}&((\sigma^2)^\star,\{b_k\},\{\bm{\mathrm{l}}_k\}) = - LM \ln(LM) + \\
& L M \ln \left(\sum_{l=1}^L \|\bm{y}_l - \bm{A} \operatorname{diag}(\bm{b}) \bm{V}^T \bm{x}_l\|^2\right) + LM.
\end{align}
Ignoring the first and the third constant terms and noting that the natural logarithm function is monotonically increasing, the minimization of $f_1^{\text{WGN}}$ in (\ref{equ:neg_likelihood_with_sigma2}) is equivalent to minimizing
%
\begin{align}\label{eq:f_2_wgn}
& f_2^{\text{WGN}}(\bm{b},\{\bm{\mathrm{l}}_k\})  = \sum_{l=1}^L \|\bm{y}_l - \bm{A} \operatorname{diag}(\bm{b}) \bm{V}^T \bm{x}_l\|^2 \\
\nonumber
& = \sum_{l=1}^L \|(\bm{y}_l - \sum_{k \neq i} b_k \bm{a}(\bm{\mathrm{l}}_k) \bm{v}^T(\bm{\mathrm{l}}_k) \bm{x}_l) - b_i \bm{a}(\bm{\mathrm{l}}_i) \bm{v}^T(\bm{\mathrm{l}}_i) \bm{x}_l \|^2.
\end{align}

Taking the derivative with respect to $b_i^*$ and setting it equal to zero, i.e., $\frac{\partial f_2^{\text{WGN}}}{\partial b_i^*} = 0$, we have 
\begin{align}\label{eq:b_i_opt_wgn}
b_i^{\text{opt}} = \frac{\bm{a}^H(\bm{\mathrm{l}}_i) \bm{Y}_i^{\text{opt}} \bm{X}^H \bm{v}^*(\bm{\mathrm{l}}_i)}{ \|\bm{a}(\bm{\mathrm{l}}_i)\|^2 \bm{v}^T(\bm{\mathrm{l}}_i) \bm{X} \bm{X}^H \bm{v}^*(\bm{\mathrm{l}}_i)}, \quad \forall i \in \mathcal{K},
\end{align}
where 
\begin{align}\label{eq:Y_i_opt_wgn}
\bm{Y}_i^{\text{opt}} = \bm{Y} - \sum_{k \neq i} b_k^{\text{opt}} \bm{a}(\bm{\mathrm{l}}_k)  \bm{v}^T(\bm{\mathrm{l}}_k) \bm{X}.
\end{align}
Define
\begin{align}\label{eq:lamda_def}
\lambda_i  =  \frac{\bm{a}^H(\bm{\mathrm{l}}_i) \bm{Y} \bm{X}^H \bm{v}^*(\bm{\mathrm{l}}_i)}{ \|\bm{a}(\bm{\mathrm{l}}_i)\|^2 \bm{v}^T(\bm{\mathrm{l}}_i) \bm{X} \bm{X}^H \bm{v}^*(\bm{\mathrm{l}}_i)}, 
\end{align}
\begin{align}\label{eq:beta_def}
\beta_{i,k} = \frac{\bm{a}^H(\bm{\mathrm{l}}_i) \bm{a}(\bm{\mathrm{l}}_k) \bm{v}^T(\bm{\mathrm{l}}_k) \bm{X} \bm{X}^H \bm{v}^*(\bm{\mathrm{l}}_i)}{ \|\bm{a}(\bm{\mathrm{l}}_i)\|^2 \bm{v}^T(\bm{\mathrm{l}}_i) \bm{X} \bm{X}^H \bm{v}^*(\bm{\mathrm{l}}_i)}.
\end{align}
Then, (\ref{eq:b_i_opt_wgn}) becomes
\begin{align} \label{eq:Linear_eq_wgn}
b_i^{\text{opt}} + \sum_{k \neq i} b_k^{\text{opt}} \beta_{i,k} = \lambda_i, \quad \forall i \in \mathcal{K},
\end{align}
or equivalently,
\begin{align}
\nonumber
\underbrace{\left[\begin{array}{llll}
1 & \beta_{1,2}  & \ldots & \beta_{1,K} \\
\beta_{2,1} & 1  & \ldots & \beta_{2,K} \\
\vdots & \vdots & \ddots  & \vdots \\
\beta_{K,1} & \beta_{K,2} &  \ldots & 1
\end{array}\right]}_{\bm{\Sigma}} \underbrace{\left[\begin{array}{l}
b_1^{\text{opt}} \\
b_2^{\text{opt}} \\
\vdots \\
b_K^{\text{opt}} 
\end{array}\right]}_{\bm{b}^{\text{opt}}} = \underbrace{\left[\begin{array}{l}
\lambda_1 \\
\lambda_2 \\
\vdots \\
\lambda_K 
\end{array}\right]}_{\bm{\lambda}}, 
\end{align}
where $\bm{\Sigma} \in \mathbb{C}^{K \times K}$. 
Thus, 
\begin{align}\label{eq:b_opt_sol}
\bm{b}^{\text{opt}} = \bm{\Sigma}^{-1} \bm{\lambda}.
\end{align}
Substituting (\ref{eq:b_opt_sol}) into $f_2^{\text{WGN}}(\bm{b},\{\bm{\mathrm{l}}_k\})$ in (\ref{eq:f_2_wgn}), we obtain the concentrated negative log-likelihood function
\begin{align}\label{eq:tilde_f_1}
f_3^{\text{WGN}}(\{\bm{\mathrm{l}}_k\}) = f_2^{\text{WGN}}(\bm{\Sigma}^{-1} \bm{\lambda},\{\bm{\mathrm{l}}_k\}).
\end{align}
When there exist only a single target, (\ref{eq:b_opt_sol}) is reduced to $b_1^{\text{opt}} = \lambda_1$. Based on $f_3^{\text{WGN}}(\{\bm{\mathrm{l}}_k\})$, we can apply the cyclic optimization and iterative fine grid 3D searching technique in Section \ref{subsec:alg_3D_ACO} for 3D-ACO. 
The remaining procedure of 3D-CO-WGN is same as 3D-ACO except that the minimization function is $f_3^{\text{WGN}}$ in (\ref{eq:tilde_f_1}) instead. Thus, 3D-CO-WGN is the same as Algorithm 1 except that $f_3$ is replaced by $f_3^{\text{WGN}}$.

\section{Numerical Results} \label{Section_results}

This section provides numerical results to validate the near-field CRB and evaluates the performance of the proposed two estimators for 3D multiple targets localization - 3D-ACO and 3D-CO-WGN under different setups.



\subsection{CRB Behavior Analysis}
\label{subsec:CRB_analysis}
We first demonstrate the necessity of considering distance-dependent channel amplitude variations in (\ref{equ:steer_Rx}) and (\ref{equ:steer_Tx}).
For comparison, we consider the conventional analysis without varying amplitude across antennas \cite{khamidullina2021conditional}. 
To ensure fair comparison, we normalize the constant-amplitude term, i.e., $\alpha^r(\bm{\mathrm{l}}_k)$ and $\alpha^t(\bm{\mathrm{l}}_k)$, into the complex reflection coefficient $\{b_k\}$ and accordingly get the complex reflection coefficient with round-trip path-loss, denoted as $\{\tilde{b}_k\}$.
Specifically, the CRB without considering amplitude variations is derived based on the following received signal model:
\begin{align}\label{equ:Rx_data_matrix_wo_pl}
\tilde{\bm{Y}} = \sum_{k = 1}^{K} \tilde{b}_k \tilde{\bm{a}}(\bm{\mathrm{l}}_k)  \tilde{\bm{v}}^T(\bm{\mathrm{l}}_k) \bm{X} + \bm{Z},
\end{align}
where $\left[\tilde{\bm{a}}(\bm{\mathrm{l}}_k)\right]_m = e^{-\mathrm{j} \nu \|\bm{\mathrm{l}}^r_m - \bm{\mathrm{l}}_{k}\|}, \left[\tilde{\bm{v}}(\bm{\mathrm{l}}_k)\right]_n = e^{-\mathrm{j} \nu \|\bm{\mathrm{l}}^t_n - \bm{\mathrm{l}}_{k}\|}$, and 
\begin{align}\label{eq:tilde_b_k}
\tilde{b}_k = \left(\frac{\lambda}{4 \pi}\right)^2   \frac{1}{\|\bm{\mathrm{l}}^r_o-\bm{\mathrm{l}}_{k}\|} \frac{1}{\|\bm{\mathrm{l}}^t_o-\bm{\mathrm{l}}_{k}\|} b_k, 
\end{align}
where $\bm{\mathrm{l}}^r_o$ and $\bm{\mathrm{l}}^t_o$ are the 3D coordinates of the reference point at Rx and Tx, respectively.
Here, we choose the path-loss with respect to the reference points and accordingly obtain $\{\tilde{b}_k\}$.
\begin{figure}[t]
\centering
\setlength{\abovecaptionskip}{+4mm}
\setlength{\belowcaptionskip}{+1mm}
\subfigure[Target on $(0,0,d)$.]{ \label{fig:CRB_wo_pl_analysis_00d}
\includegraphics[width=1.6735in]{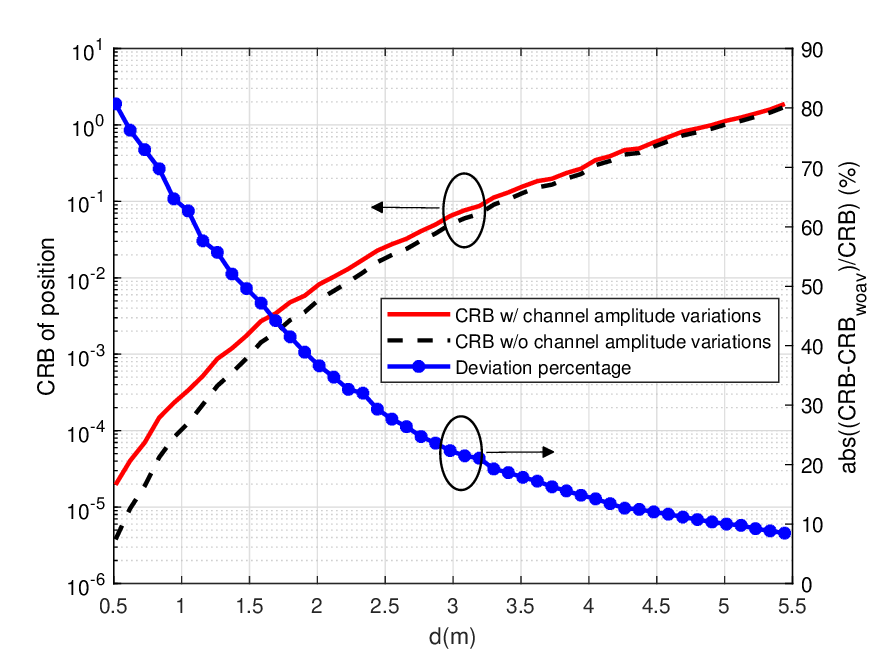}}
\subfigure[Target on $(0,6d_\mathrm{y},d_\mathrm{y})$ with $d = \sqrt{37} d_\mathrm{y}$.]{ \label{fig:CRB_wo_pl_analysis_06dd}
\includegraphics[width=1.6735in]{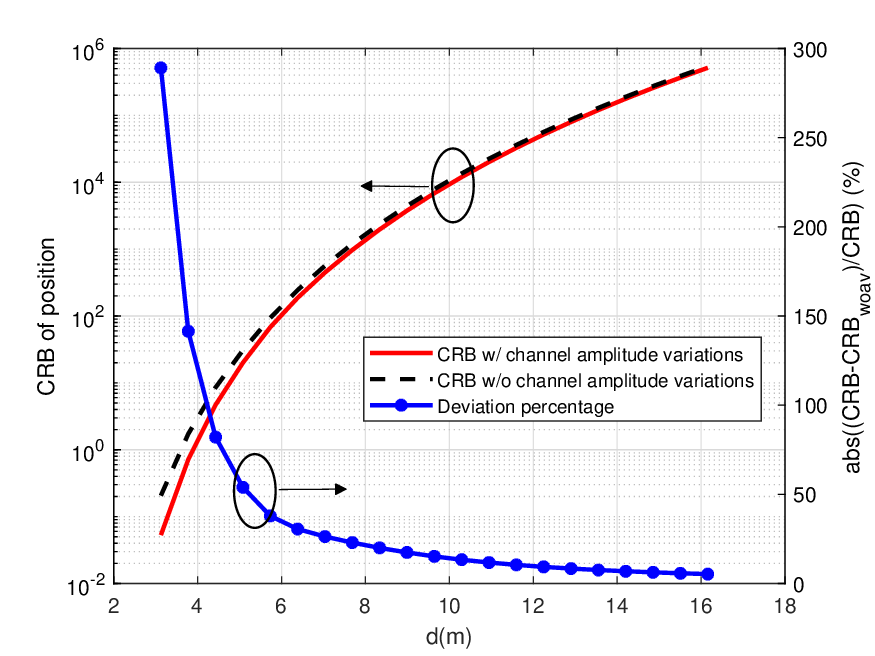}}
\caption{The comparison of the position CRB in (\ref{eq:CRB_position}) under our considered setup versus that without channel amplitude variations across antennas.}
\label{fig:CRB_wo_pl_analysis}
\end{figure}

Consider a monostatic MIMO radar setup similar to the one in Fig. \ref{fig:sm_sim}, in which both Rx and Tx are completely overlapping. Accordingly, we assume $\bm{\mathrm{l}}^r_o = \bm{\mathrm{l}}^t_o = \left[0,0,0\right]$.
Without loss of generality, we further consider a UPA in a rectangular shape with total antenna number $M = N = n_\mathrm{x} \times n_\mathrm{y} = 16 \times 768 = 12288$, where $n_\mathrm{x}$ and $n_\mathrm{y}$ are the number of the antenna along the $\mathrm{x}$-axis and the $\mathrm{y}$-axis, respectively. 
The spacing between adjacent antennas is half-wavelength with $f_c = 28$ GHz. Accordingly, the size of the array is roughly $0.1$ m $\times$ $4$ m. We further consider the simple WGN model, i.e., $\bm{Q} = \sigma^2 \bm{I}$ with $\sigma^2 = 0.001$, and adopt an isotropic transmission with unit power, i.e., each column of $\bm{X}$ is generated according to $\bm{x}_l \sim \mathcal{CN}(\bm{0},\bm{I})$ with $L = 256$. 

We first assume that there exists a target lying on the $\mathrm{z}$-axis with coordinate $(0,0,d)$. As shown in Fig. \ref{fig:CRB_wo_pl_analysis_00d}, when the target is close to the UPA, the CRB without amplitude variations significantly deviates from our derived CRB with amplitude variations (with $d \approx 3$m, the deviation is around 20\%). It is also observed that the CRB without varying amplitudes is actually smaller than that with varying amplitudes under this setup. This is because when normalizing the path-loss with respect to the center of the array in (\ref{eq:tilde_b_k}), we implicitly reduce the path-loss between the target and all the non-center antennas, thus leading to smaller CRB. The considered setup in Fig. \ref{fig:CRB_wo_pl_analysis_00d} is symmetric and does not fully manifest the importance of considering amplitude variations. Thus, we further assume that the target's coordinate is $(0, 6d_\mathrm{y}, d_\mathrm{y})$ and its distance from the center of the array $d = \sqrt{(6d_\mathrm{y})^2+d_\mathrm{y}^2} = \sqrt{37} d_\mathrm{y}$. Under this setup, Fig. \ref{fig:CRB_wo_pl_analysis_06dd} shows that even when $d \approx 12$m, there still exists 10\% deviation between the two modelings in terms of CRB. We also observe that the CRB without varying amplitudes is now larger, as during the normalization in (\ref{eq:tilde_b_k}), we implicitly reduce the path-loss between the target and the antennas located in the further half of the array but also increase the path-loss between the target and the antennas in the closer half of the array, leading to larger CRB ultimately.
From Fig. \ref{fig:CRB_wo_pl_analysis}, it is necessary to consider amplitude variations in channel modeling. We also observe that when the target is moving away from the UPA, the two different models will converge. This shows that the simplified model without  amplitude variations is valid as long as the target is distant enough from the transceiver. 
\begin{figure}[t]
\centering
\includegraphics[width=1.8in]{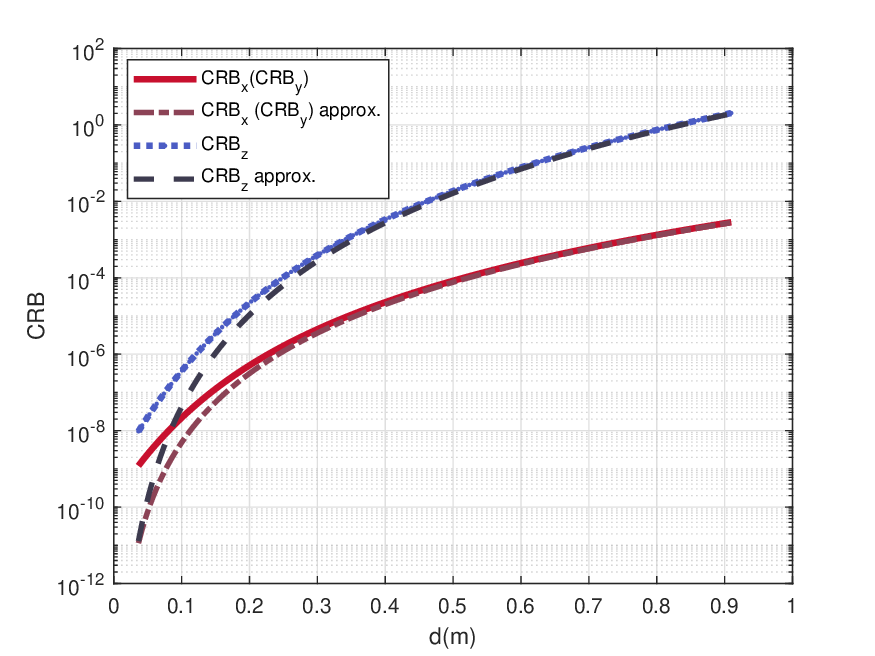}
\centering
\caption{The CRB versus the approximated expression: $\text{CRB}_\mathrm{x}$ ($\text{CRB}_\mathrm{y}$) versus the approximated $\text{CRB}_\mathrm{x}$ ($\text{CRB}_\mathrm{y}$) in (\ref{eq:CRB_x_appro_d_large}); $\text{CRB}_\mathrm{z}$ versus the approximated $\text{CRB}_\mathrm{z}$ in (\ref{eq:CRB_z_appro_d_large}).}
\label{fig:CRB_appro_d_large}
\end{figure}

We then verify the asymptotic analysis of CRB in Section \ref{subsec:closed_form_CRB_single}. Towards this end, the setup is same as above except that UPA is assumed to be a square shape with $M = N = n^2 = 35^2 = 1225$ and the target's coordinate is $(0,0,d)$. Furthermore, $\bm{R}_X = \bm{I}$. As one can observe from Fig. \ref{fig:CRB_appro_d_large}, under $f_c = 28$ GHz, when $d$ is larger than $0.3$m, the exact CRB of $\mathrm{x}$-coordinate, $\mathrm{y}$-coordinate, and $\mathrm{z}$-coordinate can be approximated relatively well by (\ref{eq:CRB_x_appro_d_large}) and (\ref{eq:CRB_z_appro_d_large}), respectively. Specifically, $\text{CRB}_\mathrm{z}$ increases much faster with respect to distance $d$ than that of $\text{CRB}_\mathrm{x}$ ($\text{CRB}_\mathrm{y}$) in Fig. \ref{fig:CRB_appro_d_large}.


\subsection{Proposed Estimators Evaluation with the CRB}


We proceed to evaluate the performance of the proposed two estimators 3D-ACO and 3D-CO-WGN with the derived CRB in terms of empirical position estimation mean square error (MSE). 
The threshold $\epsilon$ is set to $10^{-5}$ in Algorithm 1. All the curves of the estimators are obtained by averaging over 100 Monte Carol simulations.
\begin{figure}[t]
\centering
\setlength{\abovecaptionskip}{+4mm}
\setlength{\belowcaptionskip}{+1mm}
\subfigure[Setup 1 configuration: Tx array is marked in red and Rx in blue.]{ \label{fig:two_target_scen}
\includegraphics[width=1.9in]{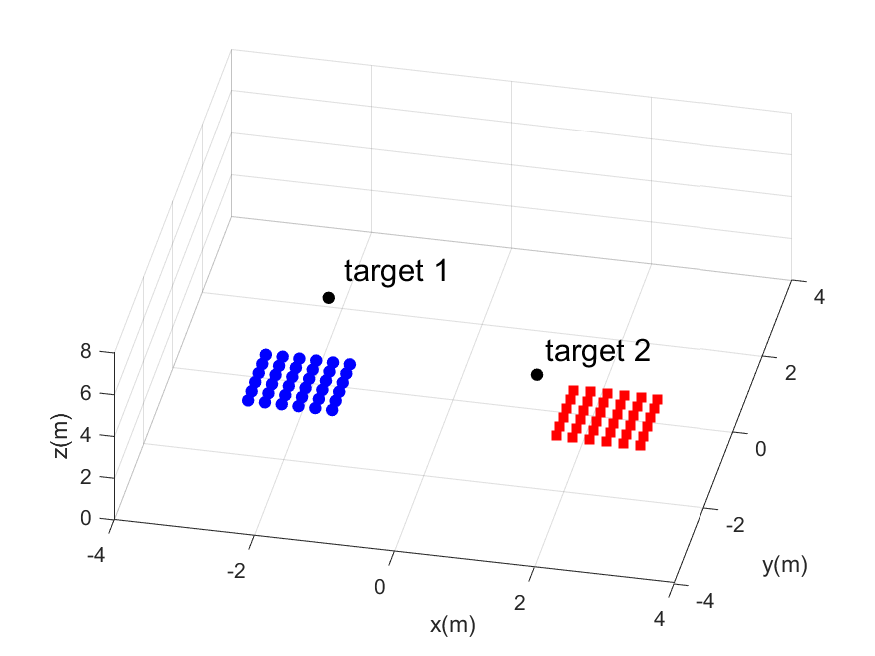}}
\subfigure[Target 1 under WGN.]{ \label{fig:MSE_1st}
\includegraphics[width=1.6735in]{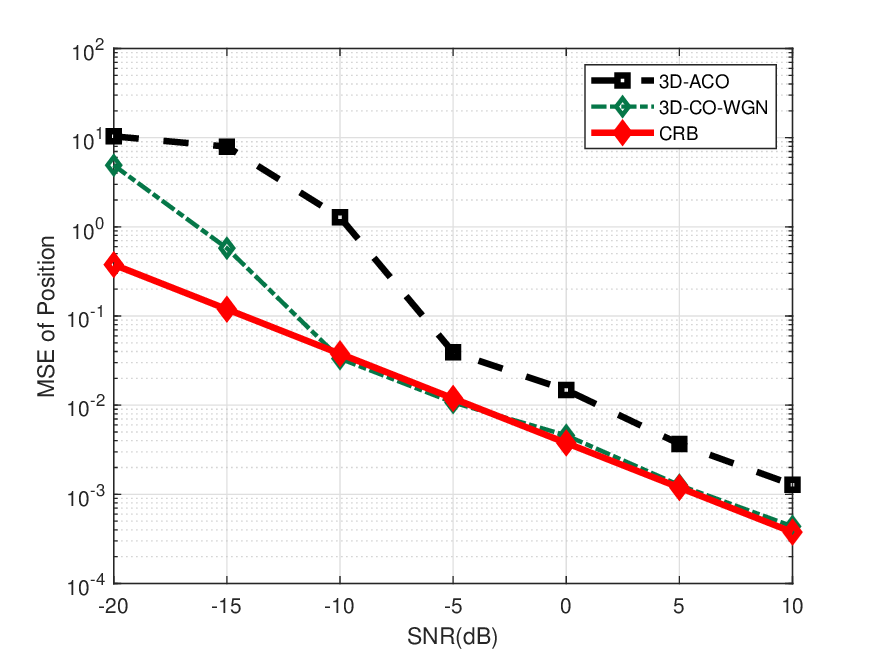}}
\subfigure[Target 2 under WGN.]{ \label{fig:MSE_2rd}
\includegraphics[width=1.6735in]{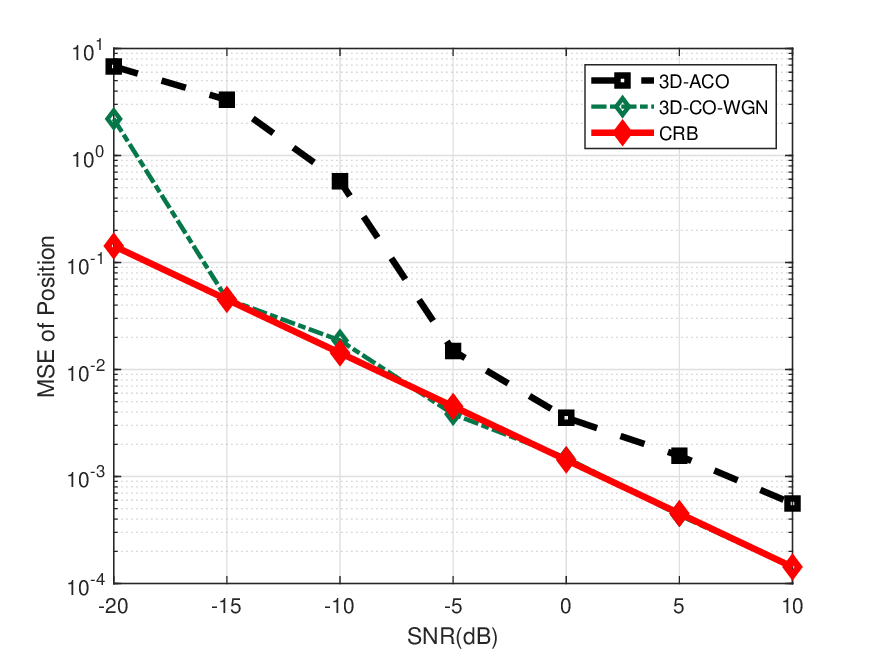}}
\subfigure[Target 1 under $\bm{Q}$ in (\ref{eq:general_Q_exp}). ]{ \label{fig:MSE_1st_general_Q}
\includegraphics[width=1.6735in]{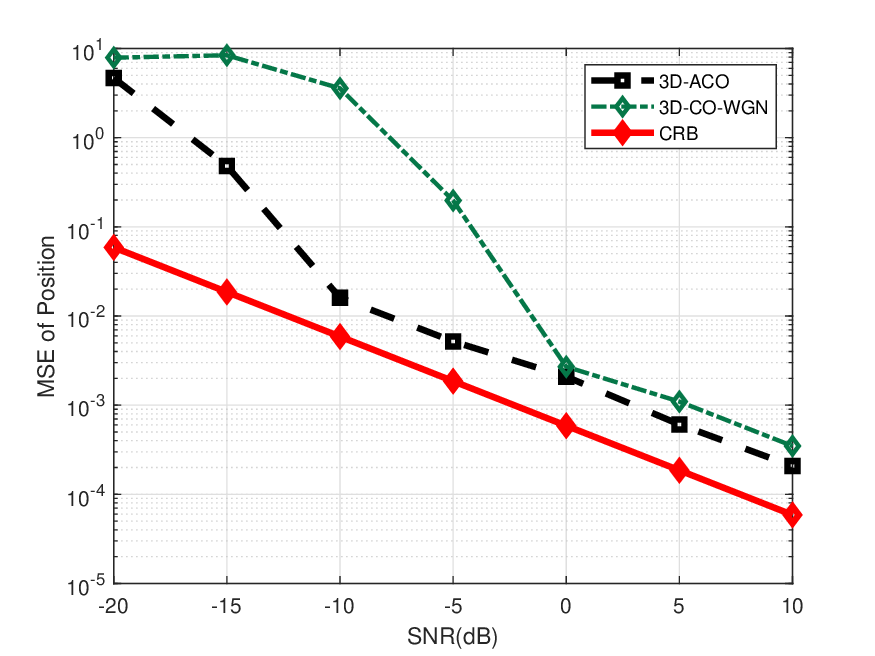}}
\subfigure[Target 2 under $\bm{Q}$ in (\ref{eq:general_Q_exp}).]{ \label{fig:MSE_2rd_general_Q}
\includegraphics[width=1.6735in]{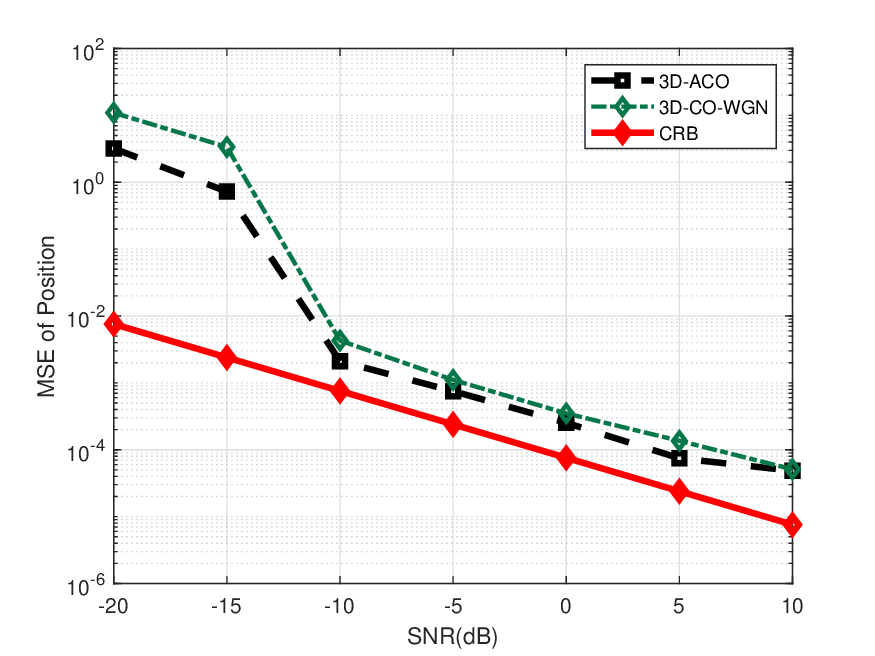}}
\caption{The performance of the proposed 3D-ACO and 3D-CO-WGN v.s. derived CRB in terms of empirical MSE with respect to SNR: (a) The considered setup; (b)-(c) The evaluation under WGN model of $\bm{Q}$; (d)-(e) The evaluation under $\bm{Q}$ specified in (\ref{eq:general_Q_exp}).}
\label{fig:multiple_nf_MSE}
\end{figure}

\begin{figure}[t]
\centering
\setlength{\abovecaptionskip}{+4mm}
\setlength{\belowcaptionskip}{+1mm}
\subfigure[Setup 2 configuration: Tx array is marked in red and Rx in blue. ]{ \label{fig:two_target_scen_setup2}
\includegraphics[width=1.9in]{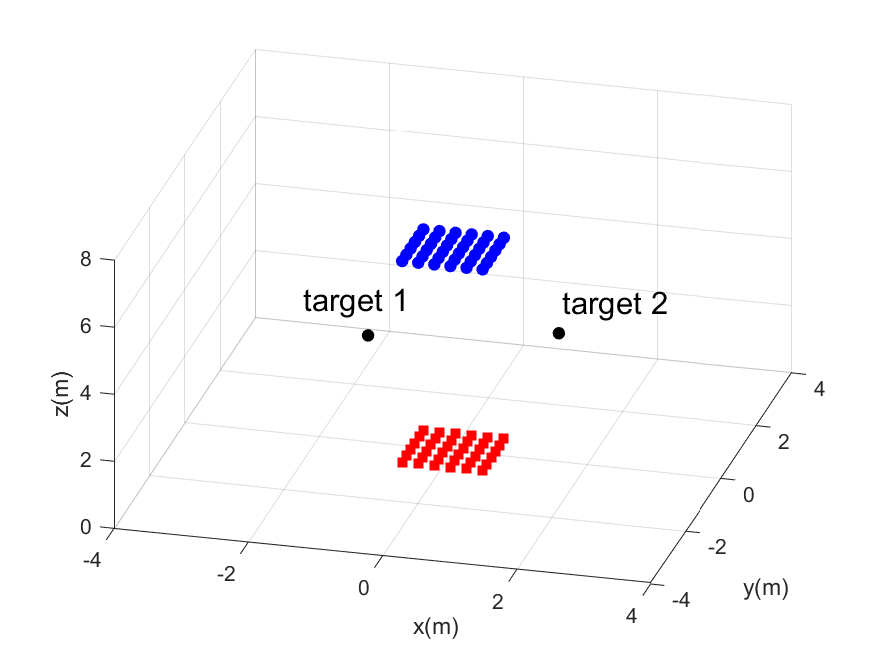}}
\subfigure[Target 1 MSE v.s. SNR. ]{ \label{fig:MSE_1st_setup2}
\includegraphics[width=1.6735in]{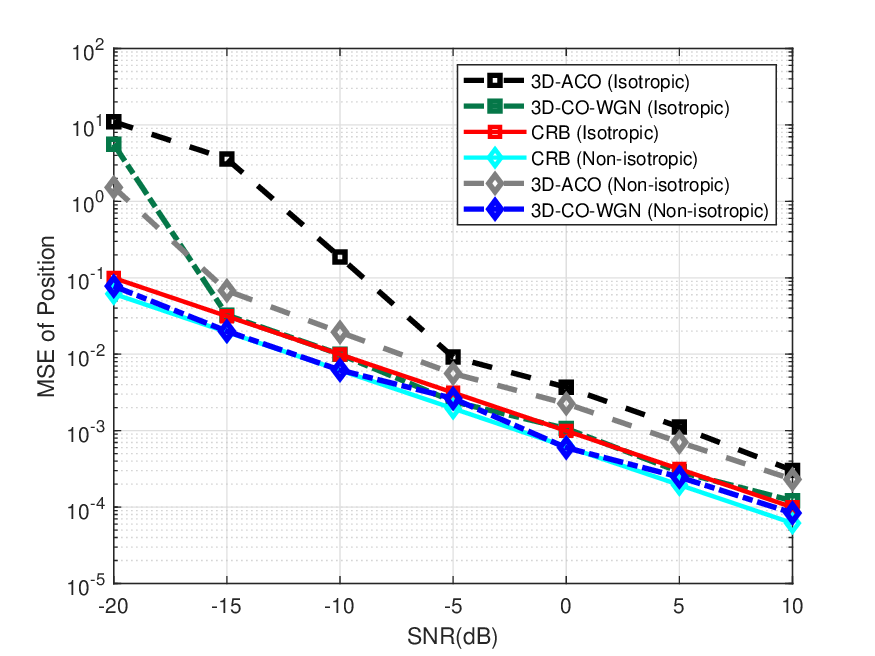}}
\subfigure[Target 2 MSE v.s. SNR.]{ \label{fig:MSE_2rd_setup2}
\includegraphics[width=1.6735in]{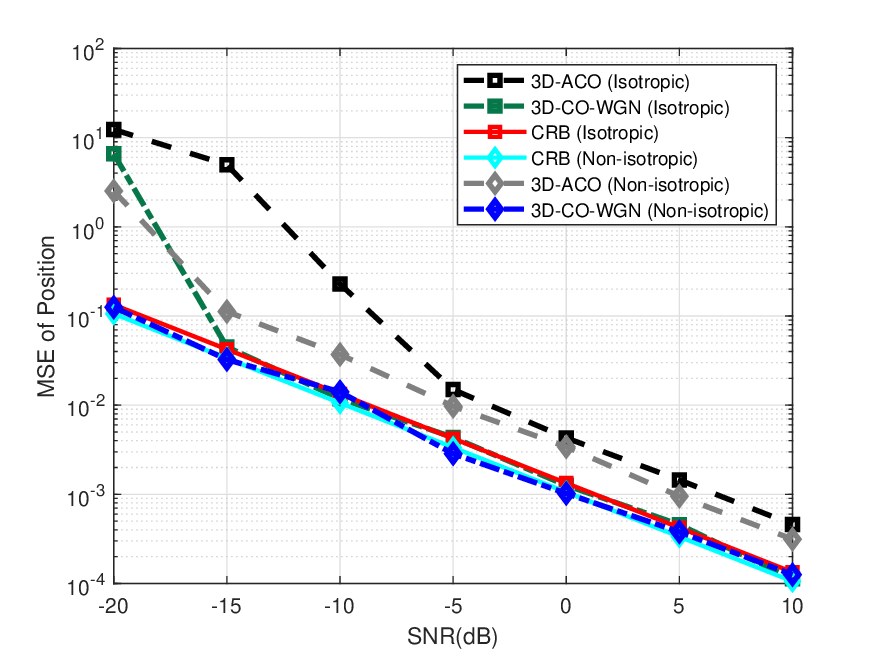}}
\caption{The performance of the proposed estimators v.s. derived CRB in terms of empirical MSE with respect to SNR under WGN model of $\bm{Q}$. Both isotropic and non-isotropic transmission strategies are considered here. }
\label{fig:multiple_nf_MSE_setup2}
\end{figure}

First, we show the impact of $\bm{Q}$ on CRB and  estimators performance. Fig. \ref{fig:two_target_scen} shows the antenna configuration corresponding to the setup depicted in Fig. \ref{fig:sm}(a) and also the relative location of the targets when there exists two targets. The center of Tx array and that of Rx array are placed at $\left[2.2, 0, 0\right]$m and $\left[-2.2, 0, 0\right]$m, respectively, while the ground truth 3D coordinates of target 1 and target 2 are $\left[-1.834, 0.294, 3.657\right]$m and $\left[1.336, -0.645, 2.898\right]$m, respectively. The complex coefficients of both targets are set to be 1. We further set $M = N = 36$, $f_c = 0.625$ GHz, and adopt the isotropic transmission as in Section \ref{subsec:CRB_analysis} with $L = 52$. The structure of $\bm{Q}$ is first chosen as $\bm{Q} = \sigma^2 \bm{I}$. In terms of CRB, Fig. \ref{fig:MSE_1st} and \ref{fig:MSE_2rd} show that although target 1 is closer to Rx array than target 2, the CRB of the position of target 2 is in general lower than that of target 1 as target 2 is much closer to Tx array than target 1. This shows that the CRB depends on the relative positions of both the targets and Tx/Rx array. In terms of estimators' performance, one observes from Fig. \ref{fig:MSE_1st} and \ref{fig:MSE_2rd} that for both targets, the performance of 3D-CO-WGN is better than 3D-ACO throughout the whole SNR regime. The underlying reason is that 3D-CO-WGN is designed based on the prior assumption of WGN model and the assumption is correct under the considered setup here. 
Next, under the same setup as in Fig. \ref{fig:two_target_scen} and the same transmitted waveform $\bm{X}$, we  consider another case when $\bm{Q}$ is not WGN but a positive definite matrix constructed as follows,
\begin{align}\label{eq:general_Q_exp}
\bm{Q}_{p,q} = 0.95^{|p-q|} e ^{\mathrm{j}(p-q) \frac{\pi}{4}},
\end{align}
where $\bm{Q}_{p,q}$ denotes the $(p,q)$-th entry of $\bm{Q}$. In terms of CRB, the different modeling of the noise and interference effect has a profound impact on the CRB of both targets.
For estimator evaluation, Fig. \ref{fig:MSE_1st_general_Q} and Fig. \ref{fig:MSE_2rd_general_Q} show that with non-WGN $\bm{Q}$, the performance of 3D-CO-WGN deteriorates while the performance of 3D-ACO remains relatively unchanged for localizing both targets. In particular, the performance of 3D-ACO is better than that of 3D-CO-WGN throughout the whole SNR regime. As the prior knowledge about the structure of $\bm{Q}$ is no longer correct, the performance of 3D-CO-WGN degrades since it is derived based on the WGN modeling of $\bm{Q}$. By contrast, 3D-ACO can be applied to general $\bm{Q}$.

Next, we show the impact of $\bm{R}_X$ on CRB and estimators performance.  Fig. \ref{fig:two_target_scen_setup2} shows another antenna configuration, which corresponds to the setup in Fig. \ref{fig:sm}(b).
The centers of Tx array and Rx array are placed at $\left[0,0,0\right]$ and $\left[0,0,6\right]$m, respectively, while the ground truth 3D coordinates of target 1 and target 2 are the same as that in Fig. \ref{fig:two_target_scen}. 
The complex coefficients of both targets are set to be 1. 
Similarly, we set $L = 52, M = N = 36$, $f_c = 0.625$ GHz, adopt the WGN model of $\bm{Q}$. For the purpose of illustration, we consider both conventional isotropic transmission with $\bm{x}_l \sim \mathcal{CN}(\bm{0},\bm{I})$ and alternative non-isotropic transmission strategies\footnote{It is possible to further optimize $\bm{R}_X$ and accordingly the waveforms to further minimize the CRBs \cite{hua2022mimojournal}. However, this problem is non-trivial in general, which is thus left for future investigation.}, in which we intuitively set $\bm{x}_l^\text{non} \sim \mathcal{CN}(\bm{0},\bm{R}_X^\text{non})$, where $\bm{R}_X^\text{non} = \frac{1}{2} \bm{I} + \frac{1}{2} \bm{t}_1 \bm{t}_1^H + \frac{1}{2} \bm{t}_2 \bm{t}_2^H$. Here, $\bm{t}_1 = \sqrt{\frac{3M}{4}} \frac{\bm{v}^*(\br{l}_1)}{\|\bm{v}(\br{l}_1)\|}$, and $\bm{t}_2 = \sqrt{\frac{M}{4}} \frac{\bm{v}^*(\br{l}_2)}{\|\bm{v}(\br{l}_2)\|}$, where $\br{l}_1$ and $\br{l}_2$ correspond to the position of target 1 and target 2, respectively. Notice that the transmitter has to know $\br{l}_1$ and $\br{l}_2$ to design $\bm{R}_X^\text{non}$, which is valid in target tracking scenarios \cite{liu2021cramer}. Note that $\operatorname{tr}(\bm{R}_X^\text{non}) = \operatorname{tr}(\bm{I}) = M$, i.e., the two strategies have the same transmit power for fair comparison.
Under the setup, Fig. \ref{fig:MSE_1st_setup2} and Fig. \ref{fig:MSE_2rd_setup2} show that compared with isotropic transmission, the considered non-isotropic one leads to smaller CRB for both targets under the same sum-power constraint. Besides, it is also observed that under the non-isotropic transmission, the performance of both estimators are generally better than that with isotropic transmission throughout the whole SNR regime.

\begin{figure}[t]
\centering
\setlength{\abovecaptionskip}{+4mm}
\setlength{\belowcaptionskip}{+1mm}
\subfigure[Setup 2 configuration with rectangular array: Tx array is marked in red and Rx in blue. ]{ \label{fig:two_target_scen_setup2_wopl}
\includegraphics[width=1.9in]{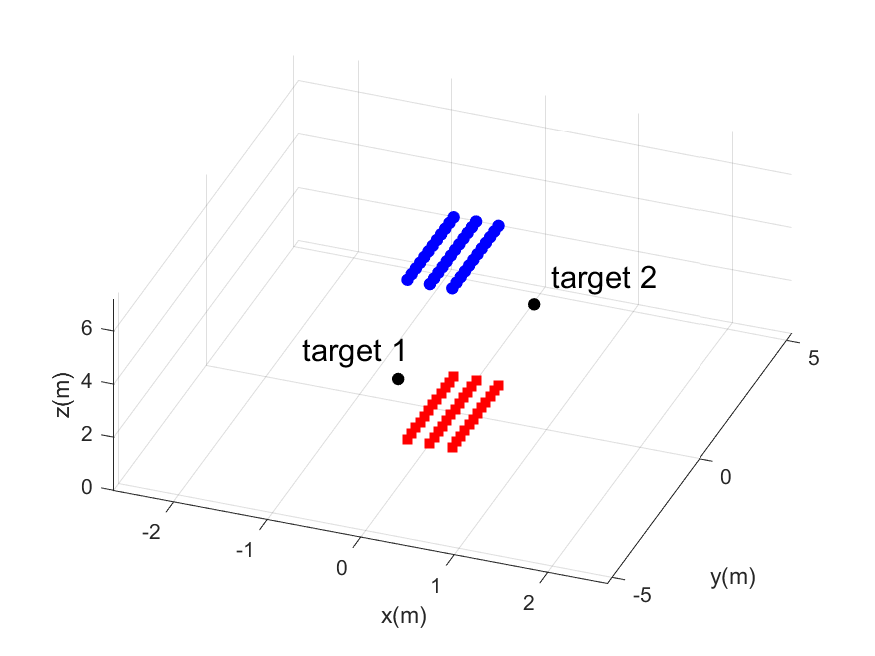}}
\subfigure[Target 1 MSE v.s. SNR. ]{ \label{fig:MSE_1st_setup2_wopl}
\includegraphics[width=1.6735in]{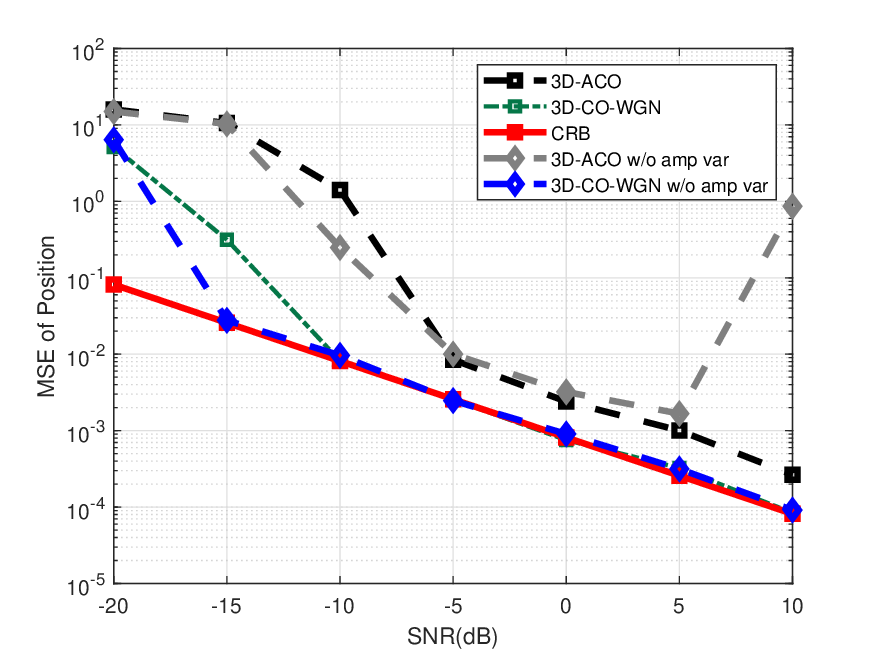}}
\subfigure[Target 2 MSE v.s. SNR.]{ \label{fig:MSE_2rd_setup2_wopl}
\includegraphics[width=1.6735in]{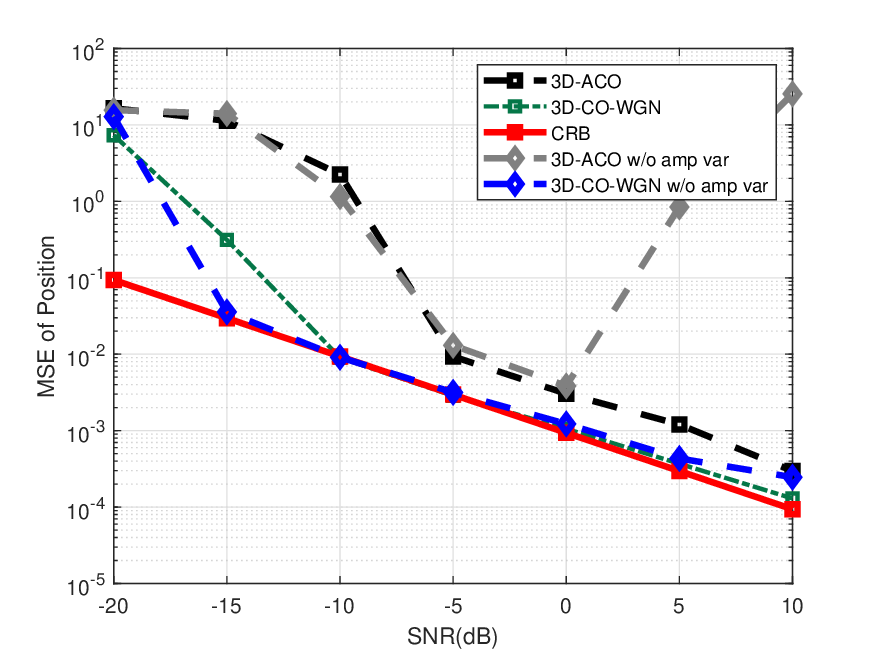}}
\caption{The performance of the proposed 3D-ACO and 3D-CO-WGN v.s. derived CRB in terms of empirical MSE with respect to SNR under WGN model of $\bm{Q}$. Here, we also consider 3D-ACO and 3D-CO-WGN without considering varying channel amplitudes.}
\label{fig:multiple_nf_MSE_setup2_wopl}
\end{figure}

Finally, we show the necessity of considering channel amplitude variations across antennas in the estimators' design. Consider the setup in Fig. \ref{fig:two_target_scen_setup2_wopl}, where we have Tx rectangular array as well as Rx rectangular array with $M = N = 3 \times 12 = 36$. The first target is located at $\left[-0.100, -2.600, 3.500\right]$m while the second target is at $\left[0.390, 2.540, 2.050\right]$m. The isotropic transmission and WGN model of $\bm{Q}$ are adopted and the other setups are same as those for Fig. \ref{fig:multiple_nf_MSE}. The received data matrix $\bm{Y}$ is still based on the model with both phase and amplitude variations but we compare the case with and without considering varying channel amplitudes when applying the two estimators to process $\bm{Y}$. As one sees, if we ignore the amplitude variations in the proposed two estimators, we find in both Fig. \ref{fig:MSE_1st_setup2_wopl} and Fig. \ref{fig:MSE_2rd_setup2_wopl} that the behaviors of the estimators will be different. In particular, without considering varying channel amplitudes, the estimators' performance of both 3D-ACO and 3D-CO-WGN degrade under high SNR regime. The 3D-ACO actually fails to work under high SNR regime, as without considering varying amplitudes, the landscape of $f_3$ in Algorithm 1 is altered, leading to incorrect localization. This shows the importance of considering complete modeling not only in CRB analysis, but in estimator design as well.


\section{Conclusion}
This paper studied a near-field MIMO radar system for multi-target localization by considering the exact spherical wavefront model with both channel phase and amplitude variations across antennas. Under this setup, we derived the CRB for estimating the 3D target locations, which is applicable in the scenario with general transmit waveforms and general noise and interference covariance matrix. We obtain the CRB in an analytical non-matrix form for the special single-target scenario and accordingly analyze its asymptotic behaviors with respect to the target distance and antenna size of the array. Next, we developed two practical localization algorithms, namely 3D-ACO and 3D-CO-WGN estimators based on the ML criterion. Numerical results showed that the consideration of antenna-specific channel amplitude variations achieves more accurate CRB and localization than that without such consideration, particularly when the target is close to the transceivers and validated the corresponding asymptotic CRB analysis. It was also shown that the performance of the proposed 3D-ACO and 3D-CO-WGN estimators approaches the CRB under different cluttered environments. Finally, the potential of non-isotropic transmit waveform has also been shown, which is expected to facilitate adaptive transmit design in near-field MIMO sensing and MIMO ISAC for future research.

\appendix

\subsection{Proof of Proposition \ref{Pro:overall_fisher}}\label{Appendix:Proof_Prop_1}

The received vectorized data matrix $\bm{y} \sim \mathcal{CN}(\bm{\mu}(\tilde{\bm{\theta}}),\bm{C}(\tilde{\bm{\theta}}))$, with $\bm{\mu}(\tilde{\bm{\theta}})$ and $\bm{C}(\tilde{\bm{\theta}})$ given in (\ref{equ:Rx_data_mat_DGC}) and (\ref{equ:cov_C}), respectively.
According to \cite[B.3.25]{stoica2005spectral},
\begin{align}\label{equ:Fisher_tilde_F_deri}
\nonumber
\tilde{\bm{\mathrm{F}}}(\tilde{\bm{\theta}}_i,\tilde{\bm{\theta}}_j)  & = \operatorname{tr}\left[\bm{C}^{-1}(\tilde{\bm{\theta}}) \frac{\partial \bm{C}(\tilde{\bm{\theta}})}{\partial \tilde{\bm{\theta}}_i} \bm{C}^{-1}(\tilde{\bm{\theta}}) \frac{\partial \bm{C}(\tilde{\bm{\theta}})}{\partial \tilde{\bm{\theta}}_j}\right]   \\
& + 2 \mathfrak{R}\left[  (\frac{\partial \bm{\mu}}{\partial \tilde{\bm{\theta}}_i})^H \bm{C}^{-1}(\tilde{\bm{\theta}}) \frac{\partial \bm{\mu}}{\partial \tilde{\bm{\theta}}_j}\right].
\end{align}
As $\bm{C}(\tilde{\bm{\theta}})$ is a block diagonal matrix, the inverse of $\bm{C}(\tilde{\bm{\theta}})$ can be obtained via the inverse of each individual block matrix $\bm{Q}$ in $\bm{C}(\tilde{\bm{\theta}})$. Thus, the first term in (\ref{equ:Fisher_tilde_F_deri}) is rewritten as
\begin{align}
\nonumber
\operatorname{tr}\left\{ \left[\begin{array}{ccc}
\bm{Q}^{-1} \frac{\partial \bm{Q}}{\partial \tilde{\bm{\theta}}_i} \bm{Q}^{-1} \frac{\partial \bm{Q}}{\partial \tilde{\bm{\theta}}_j}  & \bm{0} & \bm{0} \\
\bm{0}&  \ddots & \bm{0}\\
\bm{0}  &\bm{0} & \bm{Q}^{-1} \frac{\partial \bm{Q}}{\partial \tilde{\bm{\theta}}_i} \bm{Q}^{-1} \frac{\partial \bm{Q}}{\partial \tilde{\bm{\theta}}_j} 
\end{array}\right]  \right\},
\end{align}
which corresponds to the first term $L \operatorname{tr}\left[\bm{Q}^{-1} \frac{\partial \bm{Q}}{\partial \tilde{\bm{\theta}}_i} \bm{Q}^{-1} \frac{\partial \bm{Q}}{\partial \tilde{\bm{\theta}}_j}\right]$ in (\ref{eq:overall_fisher}). The second term in (\ref{equ:Fisher_tilde_F_deri}) is rewritten as
\begin{align}\label{equ:Gauss_CRB_sim}
& 2 \mathfrak{R} \left\{ \sum_{l=1}^{L} (\frac{\partial (\bm{A} \bm{B} \bm{V}^T \bm{x}_l)}{\partial \tilde{\bm{\theta}}_i})^H \bm{Q}^{-1} (\frac{\partial (\bm{A} \bm{B} \bm{V}^T \bm{x}_l)}{\partial \tilde{\bm{\theta}}_j}) \right\} \\
= & 2 \mathfrak{R} \operatorname{tr} \left[(\frac{\partial (\bm{A} \bm{B} \bm{V}^T \bm{X})}{\partial \tilde{\bm{\theta}}_i})^H \bm{Q}^{-1} (\frac{\partial (\bm{A} \bm{B} \bm{V}^T \bm{X})}{\partial \tilde{\bm{\theta}}_j}) \right],
\end{align}
which is the second term in (\ref{eq:overall_fisher}), completing the proof.

\subsection{Proof of Proposition \ref{Pro:F_deri}}\label{Appendix:Proof_Prop_F_deri}
As $\bm{\theta}$ in (\ref{eq:theta_interested}) does not contain any unknowns in $\bm{Q}$, the first term in (\ref{eq:overall_fisher}) vanishes. Thus, we have
\begin{align}
\nonumber
\bm{\mathrm{F}}  (\bm{\theta}_i,\bm{\theta}_j)   = 2 \mathfrak{R} \operatorname{tr} \left[ (\frac{\partial (\bm{A} \bm{B} \bm{V}^T \bm{X})}{\partial \bm{\theta}_i})^H \bm{Q}^{-1} (\frac{\partial (\bm{A} \bm{B} \bm{V}^T \bm{X})}{\partial \bm{\theta}_j}) \right].
\end{align} 

The overall structure of the FIM is given below, in which 25 blocks exist in total: 
\begin{align}
\nonumber
\left[\begin{array}{rrrrr}
\bm{\mathrm{F}}(\bm{\mathrm{x}},\bm{\mathrm{x}}) & \bm{\mathrm{F}}(\bm{\mathrm{x}},\bm{\mathrm{y}})& \bm{\mathrm{F}}(\bm{\mathrm{x}},\bm{\mathrm{z}})& \bm{\mathrm{F}}(\bm{\mathrm{x}},\bm{b}_\text{R})&\bm{\mathrm{F}}(\bm{\mathrm{x}},\bm{b}_\text{I})\\
\bm{\mathrm{F}}(\bm{\mathrm{y}},\bm{\mathrm{x}}) & \bm{\mathrm{F}}(\bm{\mathrm{y}},\bm{\mathrm{y}})& \bm{\mathrm{F}}(\bm{\mathrm{y}},\bm{\mathrm{z}})& \bm{\mathrm{F}}(\bm{\mathrm{y}},\bm{b}_\text{R})&\bm{\mathrm{F}}(\bm{\mathrm{y}},\bm{b}_\text{I})\\
\bm{\mathrm{F}}(\bm{\mathrm{z}},\bm{\mathrm{x}}) & \bm{\mathrm{F}}(\bm{\mathrm{z}},\bm{\mathrm{y}})& \bm{\mathrm{F}}(\bm{\mathrm{z}},\bm{\mathrm{z}})& \bm{\mathrm{F}}(\bm{\mathrm{z}},\bm{b}_\text{R})&\bm{\mathrm{F}}(\bm{\mathrm{z}},\bm{b}_\text{I}) \\
\bm{\mathrm{F}}(\bm{b}_\text{R},\bm{\mathrm{x}}) & \bm{\mathrm{F}}(\bm{b}_\text{R},\bm{\mathrm{y}})& \bm{\mathrm{F}}(\bm{b}_\text{R},\bm{\mathrm{z}})& \bm{\mathrm{F}}(\bm{b}_\text{R},\bm{b}_\text{R})&\bm{\mathrm{F}}(\bm{b}_\text{R},\bm{b}_\text{I})\\
\bm{\mathrm{F}}(\bm{b}_\text{I},\bm{\mathrm{x}}) & \bm{\mathrm{F}}(\bm{b}_\text{I},\bm{\mathrm{y}})& \bm{\mathrm{F}}(\bm{b}_\text{I},\bm{\mathrm{z}})& \bm{\mathrm{F}}(\bm{b}_\text{I},\bm{b}_\text{R})&\bm{\mathrm{F}}(\bm{b}_\text{I},\bm{b}_\text{I})\\
\end{array}\right]
\end{align}
However, the FIM is always symmetric positive semidefinite, and thus, we reduce the number of blocks needed to check.

First, we find $\bm{\mathrm{F}}(\bm{\mathrm{x}},\bm{\mathrm{x}})$. Towards this end, we have 
\begin{align}
\nonumber
\bm{\mathrm{F}}(\mathrm{x}_i,\mathrm{x}_j) = 2 \mathfrak{R} \operatorname{tr} \left[ (\frac{\partial (\bm{A} \bm{B} \bm{V}^T \bm{X})}{\partial \mathrm{x}_i})^H \bm{Q}^{-1} (\frac{\partial (\bm{A} \bm{B} \bm{V}^T \bm{X})}{\partial \mathrm{x}_j}) \right].
\end{align}
Notice that for arbitrary matrices $\bm{A}$ and $\bm{B}$ depending on parameter $\rho$, we have
\begin{align} \label{equ:mat_der}
\frac{\partial (\bm{A}\bm{B})}{\partial \rho} = \frac{\partial \bm{A}}{\partial \rho} \bm{B} + \bm{A} \frac{\partial \bm{B}}{\partial \rho}.
\end{align}
Thus, 
\begin{align}\label{equ:Gauss_CRB_x_left}
\frac{\partial\left(\bm{A B} \bm{V}^T \boldsymbol{X}\right)}{\partial \mathrm{x}_i}=\dot{\bm{A}_{\bm{\mathrm{x}}}} \bm{e}_i \bm{e}_i^T \bm{B} \bm{V}^T \boldsymbol{X}+\bm{A B} \bm{e}_i \bm{e}_i^T \dot{\bm{V}}_{\bm{\mathrm{x}}}^T \boldsymbol{X},
\end{align}
where $\bm{e}_i$ denotes the $i$-th column of the identity matrix. Then 
\begin{align}\label{equ:Gauss_CRB_x1}
\nonumber
\bm{\mathrm{F}}(\mathrm{x}_i,&\mathrm{x}_j)   = 2 \mathfrak{R} \operatorname{tr} \left[ \left(\dot{\bm{A}_{\bm{\mathrm{x}}}} \bm{e}_i \bm{e}_i^T \bm{B} \bm{V}^T \boldsymbol{X}+\bm{A B} \bm{e}_i \bm{e}_i^T \dot{\bm{V}}_{\bm{\mathrm{x}}}^T \boldsymbol{X}\right)^H \right.\\
& \left. \bm{Q}^{-1} \left(\dot{\bm{A}_{\bm{\mathrm{x}}}} \bm{e}_j \bm{e}_j^T \bm{B} \bm{V}^T \boldsymbol{X}+\bm{A B} \bm{e}_j \bm{e}_j^T \dot{\bm{V}}_{\bm{\mathrm{x}}}^T \boldsymbol{X}\right) \right].
\end{align}
Now,
\begin{align}
\nonumber
\operatorname{tr} & \left[ \left(\dot{\bm{A}_{\bm{\mathrm{x}}}} \bm{e}_i \bm{e}_i^T \bm{B} \bm{V}^T \boldsymbol{X}\right)^H \bm{Q}^{-1} \left(\dot{\bm{A}_{\bm{\mathrm{x}}}} \bm{e}_j \bm{e}_j^T \bm{B} \bm{V}^T \boldsymbol{X}\right) \right] \\
\nonumber
& = \bm{e}_i^T (\dot{\bm{A}}_{\bm{\mathrm{x}}}^H  \bm{Q}^{-1} \dot{\bm{A}_{\bm{\mathrm{x}}}}) \bm{e}_j \bm{e}_j^T (\bm{B} \bm{V}^T \bm{X} \bm{X}^H \bm{V}^* \bm{B}^H) \bm{e}_i \\
\nonumber
& = L (\dot{\bm{A}}_{\bm{\mathrm{x}}}^H  \bm{Q}^{-1} \dot{\bm{A}_{\bm{\mathrm{x}}}})_{ij} (\bm{B}^* \bm{V}^H \bm{R}_X^* \bm{V} \bm{B})_{ij},
\end{align}
which corresponds to the first term of the cross product in (\ref{equ:Gauss_CRB_x1}). Here, $\bm{M}_{ij}$ denotes the $(i,j)$-th element of $\bm{M}$. The other three terms have similar forms. Thus, 
$\bm{\mathrm{F}}(\bm{\mathrm{x}},\bm{\mathrm{x}}) = 2 \mathfrak{R}(\bm{\mathrm{F}}_{\bm{\mathrm{xx}}})$, where
\begin{align}\label{equ:Gauss_CRB_Fxx}
\nonumber
\bm{\mathrm{F}}_{\bm{\mathrm{xx}}} & = L (\dot{\bm{A}}_{\bm{\mathrm{x}}}^H  \bm{Q}^{-1} \dot{\bm{A}_{\bm{\mathrm{x}}}}) \odot (\bm{B}^* \bm{V}^H \bm{R}_X^* \bm{V} \bm{B}) \\
\nonumber
& + L (\dot{\bm{A}}_{\bm{\mathrm{x}}}^H  \bm{Q}^{-1} \bm{A}) \odot (\bm{B}^* \bm{V}^H \bm{R}_X^* \dot{\bm{V}}_{\bm{\mathrm{x}}} \bm{B}) \\
\nonumber
& + L (\bm{A}^H  \bm{Q}^{-1} \dot{\bm{A}_{\bm{\mathrm{x}}}}) \odot (\bm{B}^* \dot{\bm{V}}_{\bm{\mathrm{x}}}^H \bm{R}_X^* \bm{V} \bm{B})\\
& + L (\bm{A}^H  \bm{Q}^{-1} \bm{A}) \odot (\bm{B}^* \dot{\bm{V}}_{\bm{\mathrm{x}}}^H \bm{R}_X^* \dot{\bm{V}}_{\bm{\mathrm{x}}} \bm{B}).
\end{align}
Accordingly, $\bm{\mathrm{F}}_{\bm{\mathrm{yy}}}$ and $\bm{\mathrm{F}}_{\bm{\mathrm{zz}}}$ can be found by changing $\bm{\mathrm{x}}$ in (\ref{equ:Gauss_CRB_Fxx}) into $\bm{\mathrm{y}}$ and $\bm{\mathrm{z}}$, respectively. Accordingly, $\bm{\mathrm{F}}(\bm{\mathrm{y}},\bm{\mathrm{y}}) = 2 \mathfrak{R}(\bm{\mathrm{F}}_{\bm{\mathrm{yy}}})$, $\bm{\mathrm{F}}(\bm{\mathrm{z}},\bm{\mathrm{z}}) = 2 \mathfrak{R}(\bm{\mathrm{F}}_{\bm{\mathrm{zz}}})$.

Second, we find $\bm{\mathrm{F}}(\bm{\mathrm{x}},\bm{\mathrm{y}})$. In this regard, one can similarly derive that
\begin{align}\label{equ:Gauss_CRB_xy}
\nonumber
\bm{\mathrm{F}}(\mathrm{x}_i,&\mathrm{y}_j)
= 2 \mathfrak{R} \operatorname{tr} \left[ \left(\dot{\bm{A}_{\bm{\mathrm{x}}}} \bm{e}_i \bm{e}_i^T \bm{B} \bm{V}^T \boldsymbol{X}+\bm{A B} \bm{e}_i \bm{e}_i^T \dot{\bm{V}}_{\bm{\mathrm{x}}}^T \boldsymbol{X}\right)^H \right.\\
& \left. \bm{Q}^{-1} \left(\dot{\bm{A}_{\bm{\mathrm{y}}}} \bm{e}_j \bm{e}_j^T \bm{B} \bm{V}^T \boldsymbol{X}+\bm{A B} \bm{e}_j \bm{e}_j^T \dot{\bm{V}}_{\bm{\mathrm{y}}}^T \boldsymbol{X}\right) \right],
\end{align}
and find that $\bm{\mathrm{F}}(\bm{\mathrm{x}},\bm{\mathrm{y}}) = 2 \mathfrak{R}(\bm{\mathrm{F}}_{\bm{\mathrm{xy}}})$, where
\begin{align}\label{equ:Gauss_CRB_XY}
\nonumber
\bm{\mathrm{F}}_{\bm{\mathrm{xy}}} &= L (\dot{\bm{A}}_{\bm{\mathrm{x}}}^H  \bm{Q}^{-1} \dot{\bm{A}_{\bm{\mathrm{y}}}}) \odot (\bm{B}^* \bm{V}^H \bm{R}_X^* \bm{V} \bm{B}) \\
\nonumber
& + L (\dot{\bm{A}}_{\bm{\mathrm{x}}}^H  \bm{Q}^{-1} \bm{A}) \odot (\bm{B}^* \bm{V}^H \bm{R}_X^* \dot{\bm{V}}_{\bm{\mathrm{y}}} \bm{B}) \\
\nonumber
& + L (\bm{A}^H  \bm{Q}^{-1} \dot{\bm{A}_{\bm{\mathrm{y}}}}) \odot (\bm{B}^* \dot{\bm{V}}_{\bm{\mathrm{x}}}^H \bm{R}_X^* \bm{V} \bm{B})\\
& + L (\bm{A}^H  \bm{Q}^{-1} \bm{A}) \odot (\bm{B}^* \dot{\bm{V}}_{\bm{\mathrm{x}}}^H \bm{R}_X^* \dot{\bm{V}}_{\bm{\mathrm{y}}} \bm{B}).
\end{align}
We can find $\bm{\mathrm{F}}_{\bm{\mathrm{xz}}}$ and $\bm{\mathrm{F}}_{\bm{\mathrm{yz}}}$ by changing $\bm{\mathrm{xy}}$ in (\ref{equ:Gauss_CRB_xy}) and (\ref{equ:Gauss_CRB_XY}) into $\bm{\mathrm{xz}}$ and $\bm{\mathrm{yz}}$, respectively, and accordingly find $\bm{\mathrm{F}}(\bm{\mathrm{x}},\bm{\mathrm{z}}) = 2 \mathfrak{R}(\bm{\mathrm{F}}_{\bm{\mathrm{xz}}})$ and $\bm{\mathrm{F}}(\bm{\mathrm{y}},\bm{\mathrm{z}}) = 2 \mathfrak{R}(\bm{\mathrm{F}}_{\bm{\mathrm{yz}}})$.
As the FIM is inherently a symmetric positive semidefinite matrix, we have $\bm{\mathrm{F}}(\bm{\mathrm{y}},\bm{\mathrm{x}}) = \bm{\mathrm{F}}^T(\bm{\mathrm{x}},\bm{\mathrm{y}})$, $\bm{\mathrm{F}}(\bm{\mathrm{z}},\bm{\mathrm{x}}) = \bm{\mathrm{F}}^T(\bm{\mathrm{x}},\bm{\mathrm{z}})$, and $\bm{\mathrm{F}}(\bm{\mathrm{z}},\bm{\mathrm{y}}) = \bm{\mathrm{F}}^T(\bm{\mathrm{y}},\bm{\mathrm{z}})$.


Third, we find $\bm{\mathrm{F}}(\bm{b}_\text{R},\bm{b}_\text{R})$, $\bm{\mathrm{F}}(\bm{b}_\text{I},\bm{b}_\text{I})$, and $\bm{\mathrm{F}}(\bm{b}_\text{R}, \bm{b}_\text{I})$. Towards this end, we have
\begin{align}
\nonumber
\bm{\mathrm{F}}(b_{\text{R}_i},b_{\text{R}_j}) &= 2 \mathfrak{R}  \operatorname{tr} \left[ (\frac{\partial (\bm{A} \bm{B} \bm{V}^T \bm{X})}{\partial b_{\text{R}_i}})^H \bm{Q}^{-1} (\frac{\partial (\bm{A} \bm{B} \bm{V}^T \bm{X})}{\partial b_{\text{R}_j}}) \right],
\end{align}
\begin{align}
\nonumber
\bm{\mathrm{F}}(b_{\text{I}_i},b_{\text{I}_j}) &= 2 \mathfrak{R}  \operatorname{tr} \left[(\frac{\partial (\bm{A} \bm{B} \bm{V}^T \bm{X})}{\partial b_{\text{I}_i}})^H \bm{Q}^{-1} (\frac{\partial (\bm{A} \bm{B} \bm{V}^T \bm{X})}{\partial b_{\text{I}_j}}) \right],
\end{align}
\begin{align}
\nonumber
\bm{\mathrm{F}}(b_{\text{R}_i},b_{\text{I}_j}) &= 2 \mathfrak{R}  \operatorname{tr} \left[(\frac{\partial (\bm{A} \bm{B} \bm{V}^T \bm{X})}{\partial b_{\text{R}_i}})^H \bm{Q}^{-1} (\frac{\partial (\bm{A} \bm{B} \bm{V}^T \bm{X})}{\partial b_{\text{I}_j}}) \right],
\end{align}
and 
\begin{align}\label{equ:Gauss_CRB_b_Real}
\frac{\partial\left(\bm{A B} \bm{V}^T \boldsymbol{X}\right)}{\partial b_{\text{R}_i}}&=\bm{A} \bm{e}_i \bm{e}_i^T \bm{V}^T \boldsymbol{X}, \\
\label{equ:Gauss_CRB_b_Imag}
\frac{\partial\left(\bm{A B} \bm{V}^T \boldsymbol{X}\right)}{\partial b_{\text{I}_i}}&= \mathrm{j} \bm{A} \bm{e}_i \bm{e}_i^T \bm{V}^T \boldsymbol{X}.
\end{align}
Thus, similarly, we have $\bm{\mathrm{F}}(\bm{b}_\text{R},\bm{b}_\text{R}) = 2 \mathfrak{R}(\bm{\mathrm{F}}_{\bm{\mathrm{bb}}})$, where
\begin{align}\label{equ:Gauss_CRB_b_R}
\bm{\mathrm{F}}_{\bm{\mathrm{bb}}} = L (\bm{A}^H  \bm{Q}^{-1} \bm{A}) \odot (\bm{V}^H \bm{R}_X^* \bm{V}), 
\end{align}
$\bm{\mathrm{F}}(\bm{b}_\text{I},\bm{b}_\text{I}) = 2 \mathfrak{R}(\bm{\mathrm{F}}_{\bm{\mathrm{bb}}})$, and $\bm{\mathrm{F}}(\bm{b}_\text{R},\bm{b}_\text{I}) = - 2 \mathfrak{I}(\bm{\mathrm{F}}_{\bm{\mathrm{bb}}})$
Accordingly, we have $\bm{\mathrm{F}}(\bm{b}_\text{I},\bm{b}_\text{R}) = \bm{\mathrm{F}}^T(\bm{b}_\text{R},\bm{b}_\text{I}) = - 2 \mathfrak{I}(\bm{\mathrm{F}}^T_{\bm{\mathrm{bb}}})$.

Finally, we find $\bm{\mathrm{F}}(\bm{\mathrm{x}},\bm{b}_\text{R})$ and $\bm{\mathrm{F}}(\bm{\mathrm{x}},\bm{b}_\text{I})$. We then have 
\begin{align}
\nonumber
\bm{\mathrm{F}}(\mathrm{x}_i,b_{\text{R}_j}) & = 2 \mathfrak{R} \operatorname{tr} \left[ (\frac{\partial (\bm{A} \bm{B} \bm{V}^T \bm{X})}{\partial \mathrm{x}_i})^H \bm{Q}^{-1} (\frac{\partial (\bm{A} \bm{B} \bm{V}^T \bm{X})}{\partial b_{\text{R}_j}}) \right] \\
\nonumber
& = 2 \mathfrak{R} \operatorname{tr} \left[ \left(\dot{\bm{A}_{\bm{\mathrm{x}}}} \bm{e}_i \bm{e}_i^T \bm{B} \bm{V}^T \boldsymbol{X}+\bm{A B} \bm{e}_i \bm{e}_i^T \dot{\bm{V}}_{\bm{\mathrm{x}}}^T \boldsymbol{X}\right)^H \right.\\
& \left. \bm{Q}^{-1} \left(\bm{A} \bm{e}_j \bm{e}_j^T \bm{V}^T \boldsymbol{X}\right) \right].
\end{align}
Similarly, $\bm{\mathrm{F}}(\bm{\mathrm{x}},\bm{b}_\text{R}) = \bm{\mathrm{F}}^T(\bm{b}_\text{R},\bm{\mathrm{x}}) = 2 \mathfrak{R}(\bm{\mathrm{F}}_{\bm{\mathrm{xb}}})$, where
\begin{align}\label{equ:Gauss_CRB_b_xR}
\nonumber
\bm{\mathrm{F}}_{\bm{\mathrm{xb}}} & = L (\dot{\bm{A}}_{\bm{\mathrm{x}}}^H  \bm{Q}^{-1} \bm{A}) \odot (\bm{B}^* \bm{V}^H \bm{R}_X^* \bm{V})\\
& + L (\bm{A}^H  \bm{Q}^{-1} \bm{A}) \odot (\bm{B}^* \dot{\bm{V}}_{\bm{\mathrm{x}}}^H \bm{R}_X^* \bm{V}).
\end{align}
We can find $\bm{\mathrm{F}}_{\bm{\mathrm{yb}}}$ and $\bm{\mathrm{F}}_{\bm{\mathrm{zb}}}$ by changing $\bm{\mathrm{x}}$ in (\ref{equ:Gauss_CRB_b_xR}) into $\bm{\mathrm{y}}$ and $\bm{\mathrm{z}}$, respectively. Accordingly, we have $\bm{\mathrm{F}}(\bm{\mathrm{y}},\bm{b}_\text{R}) = \bm{\mathrm{F}}^T(\bm{b}_\text{R},\bm{\mathrm{y}}) = 2 \mathfrak{R}(\bm{\mathrm{F}}_{\bm{\mathrm{yb}}})$ and $\bm{\mathrm{F}}(\bm{\mathrm{z}},\bm{b}_\text{R}) = \bm{\mathrm{F}}^T(\bm{b}_\text{R},\bm{\mathrm{z}}) = 2 \mathfrak{R}(\bm{\mathrm{F}}_{\bm{\mathrm{zb}}})$.
%

For $\bm{\mathrm{F}}(\bm{\mathrm{x}},\bm{b}_\text{I})$, $\bm{\mathrm{F}}(\bm{\mathrm{y}},\bm{b}_\text{I})$, and $\bm{\mathrm{F}}(\bm{\mathrm{z}},\bm{b}_\text{I})$, according to  (\ref{equ:Gauss_CRB_b_Imag}), we similarly have 
\begin{align}
\bm{\mathrm{F}}(\bm{\mathrm{x}},\bm{b}_\text{I}) = \bm{\mathrm{F}}^T(\bm{b}_\text{I},\bm{\mathrm{x}}) = -2 \mathfrak{I}(\bm{\mathrm{F}}_{\bm{\mathrm{xb}}}),\\
\bm{\mathrm{F}}(\bm{\mathrm{y}},\bm{b}_\text{I}) = \bm{\mathrm{F}}^T(\bm{b}_\text{I},\bm{\mathrm{y}}) = -2 \mathfrak{I}(\bm{\mathrm{F}}_{\bm{\mathrm{yb}}}),\\
\bm{\mathrm{F}}(\bm{\mathrm{z}},\bm{b}_\text{I}) = \bm{\mathrm{F}}^T(\bm{b}_\text{I},\bm{\mathrm{z}}) = -2 \mathfrak{I}(\bm{\mathrm{F}}_{\bm{\mathrm{zb}}}).
\end{align}
Combining the above yields the proof.

\subsection{Proof of Proposition \ref{prop:CRB_single_WGN}}\label{Appendix:proof_CRB_single_WGN}

Under the case of a single target, the FIM obtained in Proposition \ref{Pro:F_deri} can be recasted as 
\begin{align}\label{eq:Schur_DF}
\bm{\mathrm{F}} =\left[\begin{array}{cc}
\bm{G} & \bm{H} \\
\bm{H}^T & \bm{R}
\end{array}\right],
\end{align}
where $\bm{G} = \bm{\mathrm{F}}([1:3],[1:3]), \bm{R} = \bm{\mathrm{F}}([4:5],[4:5])$, and $\bm{H} = \bm{\mathrm{F}}([1:3],[4:5])$. The equivalent FIM is expressed as
\begin{align}
\bm{D} = \bm{G} - \bm{H}\bm{R}^{-1}\bm{H}^T,
\end{align}
which is the Schur complement of block $\bm{R}$ of $\bm{\mathrm{F}}$. Thus, according to the Matrix Inversion Lemma\cite{horn2012matrix}, we have
\begin{align}
\bm{D}^{-1} = \left[\bm{\mathrm{F}}^{-1}\right]_{3 \times 3}.
\end{align}
According to the definition of the inverse of FIM and the formula for the inverse matrix,
\begin{align}
\text{CRB}_\mathrm{x} & = \left[\bm{\mathrm{F}}^{-1}\right]_{1,1}=\frac{1}{|\bm{D}|}\left|\begin{array}{ll}
d_{2,2} & d_{2,3} \\
d_{3,2} & d_{3,3}
\end{array}\right|,\\
\text{CRB}_\mathrm{y} & = \left[\bm{\mathrm{F}}^{-1}\right]_{2,2}=\frac{1}{|\bm{D}|}\left|\begin{array}{ll}
d_{1,1} & d_{1,3} \\
d_{3,1} & d_{3,3}
\end{array}\right|,\\
\label{eq:CRB_Schur_z}
\text{CRB}_\mathrm{z} & = \left[\bm{\mathrm{F}}^{-1}\right]_{3,3}=\frac{1}{|\bm{D}|}\left|\begin{array}{ll}
d_{1,1} & d_{1,2} \\
d_{2,1} & d_{2,2}
\end{array}\right|,
\end{align}	
where $\left[\bm{\mathrm{F}}^{-1}\right]_{i,i}$ is the $(i,i)$-th element of $\bm{\mathrm{F}}^{-1}$ and $d_{i,j}$ is the $(i,j)$-th element of $\bm{D}$, respectively.
Combining the assumption of $\bm{Q} = \sigma^2 \bm{I}$, the definition of $\bm{\mathrm{F}}$ in Proposition \ref{Pro:F_deri}, and (\ref{eq:Schur_DF})-(\ref{eq:CRB_Schur_z}) yields the proof.

\subsection{Proof of Proposition \ref{prop:CRB_sim_expression}}\label{Appendix:proof_CRB_sim_expression}

Due to the symmetry of the antenna array with respect to the target, one can verify (\ref{eq:a_norm_pds})-(\ref{eq:a_x_der}) based on their definitions, the proof is omitted for simplicity. Besides that, we have
\begin{align}
& \dot{\bm{a}}_\mathrm{x}^H \bm{a} = \bm{a}^H \dot{\bm{a}}_\mathrm{x} = \dot{\bm{v}}_\mathrm{x}^H \bm{v} = \bm{v}^H \dot{\bm{v}}_\mathrm{x} =0, \\
& \dot{\bm{a}}_\mathrm{y}^H \bm{a} = \bm{a}^H \dot{\bm{a}}_\mathrm{y} = \dot{\bm{v}}_\mathrm{y}^H \bm{v} = \bm{v}^H \dot{\bm{v}}_\mathrm{y} =0, \\
\label{eq:dxyz_ortho}
& \dot{\bm{a}}_\mathrm{x}^H \dot{\bm{a}}_\mathrm{y} = \dot{\bm{a}}_\mathrm{x}^H \dot{\bm{a}}_\mathrm{z} = \dot{\bm{a}}_\mathrm{y}^H \dot{\bm{a}}_\mathrm{z} = 0, \\
\label{eq:vxyz_ortho}
&\dot{\bm{v}}_\mathrm{x}^H \dot{\bm{v}}_\mathrm{y} = \dot{\bm{v}}_\mathrm{x}^H \dot{\bm{v}}_\mathrm{z} = \dot{\bm{v}}_\mathrm{y}^H \dot{\bm{v}}_\mathrm{z} = 0,
\end{align}
while the following equations hold,
\begin{align}
\label{eq:a_norm_pds_partial}
& \|\bm{a}\|^2   =  \|\bm{v}\|^2, \enspace \|\dot{\bm{a}}_\mathrm{x}\|^2 = \|\dot{\bm{v}}_\mathrm{x}\|^2 = \|\dot{\bm{a}}_\mathrm{y}\|^2 = \|\dot{\bm{v}}_\mathrm{y}\|^2, \\
& \|\dot{\bm{a}}_\mathrm{z}\|^2 = \|\dot{\bm{v}}_\mathrm{z}\|^2, \enspace \dot{\bm{a}}_\mathrm{z}^H \bm{a}   = \dot{\bm{v}}_\mathrm{z}^H \bm{v} =  \overline{\bm{a}^H \dot{\bm{a}}_\mathrm{z}} = \overline{\bm{v}^H \dot{\bm{v}}_\mathrm{z}}.
\end{align}
Thus, under the assumption that $\bm{R}_X = \bm{I}$, all the terms in (\ref{eq:f_uu})-(\ref{eq:f_yz_R}) are equal to zero except for
\begin{align}\label{eq:sim_f_uu_R}
\tilde{f}_{\mathrm{xx}} = \tilde{f}_{\mathrm{yy}} = 2 \|\bm{a}\|^2 \|\dot{\bm{a}}_\mathrm{x}\|^2, \enspace f_{\mathrm{zz}}^R = 4 |\dot{\bm{a}}_\mathrm{z}^H \bm{a}|^2,
\end{align}
\begin{align}\label{eq:sim_f_zz}
\tilde{f}_{\mathrm{zz}} = 2 \|\bm{a}\|^2 \|\dot{\bm{a}}_\mathrm{z}\|^2 + 2 |\dot{\bm{a}}_\mathrm{z}^H \bm{a}|^2.
\end{align}
Thus, $\tilde{\bm{D}}$ in (\ref{eq:D_tilde}) is simplified into a diagonal matrix, i.e.,
\begin{align}
\nonumber
\tilde{\bm{D}} = 2 \operatorname{diag}(\|\bm{a}\|^2 \|\dot{\bm{a}}_\mathrm{x}\|^2, \|\bm{a}\|^2 \|\dot{\bm{a}}_\mathrm{x}\|^2,  \|\bm{a}\|^2 \|\dot{\bm{a}}_\mathrm{z}\|^2 - |\dot{\bm{a}}_\mathrm{z}^H \bm{a}|^2),
\end{align}
and based on Proposition \ref{prop:CRB_single_WGN},  we accordingly have
\begin{align}
\nonumber
\text{CRB}_\mathrm{x} = \text{CRB}_\mathrm{y} & = \frac{2 \sigma^2 \|\bm{a}\|^2 \|\dot{\bm{a}}_\mathrm{x}\|^2}{|b|^2 L |\tilde{\bm{D}}|}  (\|\bm{a}\|^2 \|\dot{\bm{a}}_\mathrm{z}\|^2 - |\dot{\bm{a}}_\mathrm{z}^H \bm{a}|^2 )\\
\nonumber
& = \frac{\sigma^2}{4 |b|^2 L} \frac{1}{\|\bm{a}\|^2 \|\dot{\bm{a}}_\mathrm{x}\|^2}, 
\end{align}
\begin{align}
\nonumber
\text{CRB}_\mathrm{z}  = \frac{2 \sigma^2 \|\bm{a}\|^4 \|\dot{\bm{a}}_\mathrm{x}\|^4}{|b|^2 L |\tilde{\bm{D}}|}  = \frac{\sigma^2}{4 |b|^2 L} \frac{1}{ \|\bm{a}\|^2 \|\dot{\bm{a}}_\mathrm{z}\|^2 - |\dot{\bm{a}}_\mathrm{z}^H \bm{a}|^2 }.
\end{align}
Combining the above results yields the proof.

\subsection{Proof of Proposition \ref{prop:asy_CRB_d_large}}\label{Appendix:proof_asy_CRB_d_large}

We first prove (\ref{eq:CRB_xy_asymp}). Let $\xi = \frac{\sigma^2}{2 |b|^2 L}$ and $D_a = \frac{1}{d^2} + 4 (\sum_{i=0}^{\frac{n-1}{2}} \sum_{k=1}^{\frac{n-1}{2}} \frac{1}{d^2 + i^2s^2 + k^2 s^2})$. According to (\ref{eq:CRB_sim_expression_xy}), we have
\begin{align}
\nonumber
\lim\limits_{d \to \infty} \frac{\text{CRB}_\mathrm{x}}{d^6} = \lim\limits_{d \to \infty} \frac{\text{CRB}_\mathrm{y}}{d^6} = \lim\limits_{d \to \infty}  \frac{8 \xi \nu^4}{D_a \left(D_1^\mathrm{x} + \nu^2 D_2^\mathrm{x}\right)} \frac{1}{d^6}.
\end{align}
As $d \to \infty$, $D_1^\mathrm{x}$ will become insignificant as compared to $D_2^\mathrm{x}$ and can thus be ignored. Thus,
\begin{align}
\nonumber
& \lim\limits_{d \to \infty} \frac{\text{CRB}_\mathrm{x}}{d^6}  = \lim\limits_{d \to \infty} \frac{\text{CRB}_\mathrm{y}}{d^6} = \lim\limits_{d \to \infty}  \frac{8 \xi \nu^2}{(D_a d^2)  (D_2^\mathrm{x} d^4)} \\
\nonumber
= &  8 \xi \nu^2 \left(\lim\limits_{d \to \infty} \frac{1}{D_a d^2}\right)  \left(\lim\limits_{d \to \infty} \frac{1}{D_2^\mathrm{x} d^4}\right)  \\
= &  8 \xi \nu^2 \left(\frac{1}{n^2}\right) \left(\frac{12}{s^2 (n^4- n^2)} \right) = \frac{48 \sigma^2}{|b|^2 L}  \frac{\nu^2}{(n^6-n^4)s^2}.
\end{align}
We then prove (\ref{eq:CRB_z_asymp}), as $\lim\limits_{d \to \infty} \frac{\text{CRB}_\mathrm{z}}{d^8}$ is expressed as
\begin{align}
\lim\limits_{d \to \infty} \frac{8 \xi \nu^4}{d^{10} \left[\left(D_1^\mathrm{z} D_a - (D_2^\mathrm{z})^2\right) + \nu^2 \left( D_2^\mathrm{z} D_a - (D_3^\mathrm{z})^2 \right)\right]},
\end{align}
similarly, when $d \to \infty$, $D_1^\mathrm{z}D_a - (D_2^\mathrm{z})^2$ will become insignificant as compared to $D_2^\mathrm{z}D_a - (D_3^\mathrm{z})^2$ and can be ignored. Thus,
\begin{align}\label{eq:CRB_z_after_ignore}
\lim\limits_{d \to \infty} \frac{\text{CRB}_\mathrm{z}}{d^8} =  \frac{8 \xi \nu^2}{\lim\limits_{d \to \infty} d^{10} \left[D_2^\mathrm{z} D_a - (D_3^\mathrm{z})^2\right]}. 
\end{align}
As we have the following limit equation, 
\begin{align}
\nonumber
& \lim\limits_{d \to \infty}   \left(\frac{4d^{10}}{(d^2+a)^2(d^2+b)} + \frac{4d^{10}}{(d^2+a)(d^2+b)^2}\right) \\ 
& -  \frac{8d^{10}}{(d^2+a)^{\frac{3}{2}}(d^2 + b)^{\frac{3}{2}}}  \triangleq \lim\limits_{d \to \infty} \zeta(a,b) = (a-b)^2, 
\end{align}
where $\zeta(a,b)$ is defined accordingly for simplicity. After the cancellation of the identical terms in $D_2^\mathrm{z}D_a$ and $(D_3^\mathrm{z})^2$, 
\begin{align}
\nonumber
& \lim\limits_{d \to \infty} d^{10} \left[D_2^\mathrm{z} D_a - (D_3^\mathrm{z})^2\right] = \lim\limits_{d \to \infty} \sum_{i=0}^{\frac{n-1}{2}} \sum_{k=1}^{\frac{n-1}{2}} \zeta(0,(i^2+k^2)s^2)  \\
\nonumber
& + 4\left( \underset{j^2+m^2<i^2+k^2}{\sum_{j=0}^{\frac{n-1}{2}} \sum_{m=1}^{\frac{n-1}{2}}\sum_{i=0}^{\frac{n-1}{2}} \sum_{k=1}^{\frac{n-1}{2}}} \zeta((j^2+m^2)s^2,(i^2+k^2)s^2)\right) 
\end{align}
\begin{align}
\nonumber
& = \sum_{i=0}^{\frac{n-1}{2}} \sum_{k=1}^{\frac{n-1}{2}} (i^2+k^2)^2 s^4 \\ 
\nonumber
& + 4\left(\underset{j^2+m^2<i^2+k^2}{\sum_{j=0}^{\frac{n-1}{2}} \sum_{m=1}^{\frac{n-1}{2}}\sum_{i=0}^{\frac{n-1}{2}} \sum_{k=1}^{\frac{n-1}{2}}} [(i^2+k^2)-(j^2+m^2)]^2 s^4 \right).
\end{align}
Let $\mu = \sum\limits_{i=0}^{\frac{n-1}{2}} \sum\limits_{k=1}^{\frac{n-1}{2}} (i^2+k^2)^2$ and $\beta =\left( \sum\limits_{i=0}^{\frac{n-1}{2}} \sum\limits_{k=1}^{\frac{n-1}{2}} (i^2+k^2)\right)^2$. With some manipulation, 
\begin{align}\label{eq:CRB_z_denominator}
\nonumber
& \lim\limits_{d \to \infty} d^{10} \left[D_2^\mathrm{z} D_a - (D_3^\mathrm{z})^2\right] \\
\nonumber
= &  s^4 \left( \mu + 4 \left[(\frac{n+1}{2}\frac{n-1}{2}-1)\mu - (\beta-\mu)\right] \right) \\
= & s^4 \left[n^2 \mu - 4 \beta\right].
\end{align}
As $\mu$ and $\beta$ can be calculated as a function of $n$ as follows,
\begin{align}
\label{eq:mu_cal}
\mu  = \frac{n^2 (n^2-1) (7n^2-13)}{720}, \enspace \beta  = \frac{ (n^2-1)^2 n^4 }{576}.
\end{align}
Combining (\ref{eq:CRB_z_after_ignore}), (\ref{eq:CRB_z_denominator}), and (\ref{eq:mu_cal}) yields (\ref{eq:CRB_z_asymp}), which completes the proof.

\subsection{Proof of Proposition \ref{prop:asy_CRB_P_large}}\label{Appendix:proof_asy_CRB_P_large}

When $n$ is an positive odd number, $\|\bm{a}\|^2$, $\|\dot{\bm{a}}_\mathrm{x}\|^2$, $\|\dot{\bm{a}}_\mathrm{z}\|^2$, and $|\dot{\bm{a}}_\mathrm{z}^H \bm{a}|^2$ are all larger than 0. Thus, according to Proposition \ref{prop:CRB_sim_expression}, to show (\ref{eq:CRB_xy_P_inf}), we only have to prove $\|\bm{a}\|^2$ diverges when $n \to \infty$. Towards this end, as the array is symmetric with respect to its center, one can find that $\|\bm{a}\|^2$ is lower bounded by the following double integral,
\begin{align}\label{eq:a_norm_lower_bound}
\int\limits_{0}^{\frac{n}{2}s} \int\limits_{0}^{\frac{n}{2}s} \frac{d\omega d\gamma}{d^2+(\omega+s)^2+(\gamma+s)^2} \leq s^2 \nu^2 \|\bm{a}\|^2.
\end{align}
When $n \to \infty$, the double integral diverges and thus $\|\bm{a}\|^2$ also diverges, i.e., 
\begin{align}\label{eq:a_norm_P_inf}
\lim\limits_{n \to \infty} \|\bm{a}\|^2 = \infty.
\end{align}
Accordingly, we have (\ref{eq:CRB_xy_P_inf}). With (\ref{eq:CRB_sim_expression_z}) in Proposition \ref{prop:CRB_sim_expression} and (\ref{eq:a_norm_P_inf}), we only have to additionally prove that $|\dot{\bm{a}}_\mathrm{z}^H \bm{a}|^2$ converges when $n \to \infty$ to yield (\ref{eq:CRB_z_P_inf}). Based on (\ref{eq:a_x_der}), it is equivalent to proving that $D_2^\mathrm{z}$ and $D_3^\mathrm{z}$ both converges when $n \to \infty$. For $D_2^\mathrm{z}$, it is upper bounded as follows,
\begin{align}\label{eq:D_2z_lower_bound}
\frac{D_2^\mathrm{z}}{4} s^2 \leq  \int\limits_{-s}^{\frac{n}{2}s} \int\limits_{-s}^{\frac{n}{2}s} \frac{d\omega d\gamma}{(d^2+\omega^2+\gamma^2)^2} + \frac{s^2}{4 d^4} +  \sum_{k=1}^{\frac{n-1}{2}} \frac{s^2}{(d^2+(ks)^2)^2}.
\end{align}
One can easily check that the upper bound of $D_2^\mathrm{z}$ converges when $n \to \infty$ and thus $D_2^\mathrm{z}$ also converges. For $D_3^\mathrm{z}$, we similarly have
\begin{align}\label{eq:D_3z_lower_bound}
\frac{D_3^\mathrm{z}}{4} s^2 \leq \int\limits_{-s}^{\frac{n}{2}s} \int\limits_{-s}^{\frac{n}{2}s} \frac{d\omega d\gamma}{(d^2+\omega^2+\gamma^2)^\frac{3}{2}}  + \frac{s^2}{4 d^3} + \sum_{k=1}^{\frac{n-1}{2}} \frac{s^2}{(d^2+(ks)^2)^\frac{3}{2}},
\end{align}
the upper bound of $D_3^\mathrm{z}$ converges when $n \to \infty$ and thus $D_3^\mathrm{z}$ also converges. Combining (\ref{eq:D_2z_lower_bound}) and (\ref{eq:D_3z_lower_bound}) yields that  $|\dot{\bm{a}}_\mathrm{z}^H \bm{a}|^2$ converges when $n \to \infty$ and as a result, (\ref{eq:CRB_z_P_inf}) holds.
\bibliographystyle{ieeetran}

\bibliography{refsv2}

\end{document}